\documentclass[12pt]{article}
\usepackage{amssymb}
\usepackage{amsmath}
\usepackage[ps,dvips,matrix,arrow,frame,import,curve,color]{xy}
\makeatletter
\@addtoreset{equation}{section}
\makeatother

\def\be{\begin{equation}}
\def\ee{\end{equation}}

\def\ba{\begin{array}}
\def\ea{\end{array}}

\newcommand{\bea}{\begin{eqnarray}}
\newcommand{\eea}{\end{eqnarray}}

\def\N{$\cal N$}
\def\E {$E_{7(7)}$}

\begin{document}
\hfill{}

\begin{flushright}
\end{flushright}

\vskip 2cm

\vspace{24pt}

\begin{center}
{ \LARGE {\bf    \N=8 Supergravity on the Light Cone  }}

\vspace{24pt}

{\large  {\bf   Renata Kallosh}}

    \vspace{15pt}

{Department of Physics, Stanford University, Stanford, CA 94305}

\vspace{10pt}

\vspace{24pt}

\end{center}

\begin{abstract}

 We construct the  generating functional for the light-cone superfield amplitudes in a chiral momentum superspace. It generates the  $n$-point particle amplitudes which on shell are equivalent to the covariant ones. Based on the action depending on unconstrained  light-cone chiral scalar superfield, this functional  provides a  regular d=4 QFT  path integral derivation of the Nair-type amplitude constructions.

By performing a  Fourier transform into the  light-cone chiral coordinate superspace  we find that the quantum corrections to the superfield amplitudes with $n$ legs are non-local in transverse directions for the diagrams with the number of loops smaller than at least $n+3$.  This suggests the reason why UV infinities, which are proportional to local vertices, cannot appear at least before 7 loops  in the light-cone supergraph computations. Using the  \E \, symmetry we argue that the light-cone supergraphs predict  all-loop finiteness of d=4 \N=8 supergravity.

\end{abstract}
\newpage
\tableofcontents

\section{Introduction}

\N=8 supergravity (SG) \cite{Cremmer:1979up} has 32 supersymmetries and can be described in an  on shell \footnote{In the background field method in QFT in supergravity the meaning of the ``on shell''
is that the background field satisfies fully {\it  non-linear classical field equations}. The background method, assuming that  the quantization is performed in the background covariant gauges, takes care of gauge symmetries.
}  covariant superspace with 32 coordinates \cite{Brink:1979nt}. From the covariant  geometric torsions and curvatures one can construct, starting from the 8-loop  order, an infinite number of the super-invariants which serve as candidate counterterms \cite{Howe:1980th}, \cite{Kallosh:1980fi}. There is also a restricted set of super-invariants which can be constructed at the linearized level using a 16-dimensional sub-superspace of \N=8 SG  \cite{Kallosh:1980fi}, \cite{Howe:1981xy}, \cite{Bossard:2009sy}.  They start from the 3-loop level.

A new wave of studies of \N=8 SG started more recently in \cite{Bern:1998ug} where on the basis of the unitarity cut method \cite{Bern:1994zx} a suggestion was made in \cite{Bern:2006kd} that the theory may be UV finite at all-loop orders.
Also string theory considerations bases on pure spinors \cite{Berkovits:2006vc,Green:2006gt} suggested the onset of divergences starting from the 9-loop order or even all-loop finiteness.\footnote{There was some concern expressed in \cite{Green:2007zzb} as to whether one can use string theory to evaluate the higher loop UV properties of the four-dimensional \N=8 SG. This concern originates from the non-decoupling of the states of four-dimensional \N=8 SG from the states of string theory.}

The interest to \N=8 SG increased sharply when   it  was established in  \cite{Bern:2007hh} by the unitarity cut computations  that the 3-loop \N=8 supergravity is superfinite, see the recent review in \cite{Bern:2009kf}. One would like to understand this striking result as much as possible avoiding the direct computations in a hope to predict the situation at the higher loop level where direct computations become more and more difficult.

An interesting aspect of \N=8 SG was noticed in  \cite{Schnitzer:2007kh}:
if \N=8 SG, as \N=4 SYM (supersymmetric Yang-Mills theory), lies on a Regge trajectory, it would be consistent with the conjecture that N=8 supergravity is ultraviolet finite in perturbation theory. In   \cite{ArkaniHamed:2008gz}
 it was
conjectured that \N=8 amplitudes may be completely determined by their leading singularities and this would directly imply the perturbative finiteness of \N=8 SG. Finally, in \cite{Kallosh:2008mq} it was  suggested that the counterterms in the unconstrained light-cone superspace may not be available, which may  explain the all-loop UV finiteness  of the theory.

Since the covariant superspace is based on  superfields satisfying equations of motion, there is no way to construct the supergraph Feynman rules in the Lorentz covariant superspace. At best one can  construct the candidate counterterms.  Meanwhile, \N=8 SG in the light-cone superspace with 16 Grassmann coordinates developed in   \cite{Brink:1982pd},\cite{Brink:2008qc} has an unconstrained chiral scalar superfield, which in a chiral basis depends only on 8 Grassmann coordinates. Since the chiral light-cone superfield is off shell, there is a possibility to identify the supergraph Feynman rules in the light-cone superspace as it was done in the past for \N=4 SYM in \cite{Mandelstam:1982cb}-\cite{Belitsky:2004sc}. For \N=8 SG it may be more difficult technically; moreover, only 3- and 4-coupling vertices are known so far.  Still one can hope to learn more about the UV structure of the theory by comparing it to the information known from the covariant gauges and unitarity cut methods \footnote{The concept of ``on shell'' in the recent computations of the amplitudes means that it is an S-matrix type of computations when each external particle in the process satisfies {\it free equations of motion and only physical states propagate.}
The difference between the background field method and  the recent computation of the amplitudes is subtle, but one should keep it in mind when using the covariant counterterms with respect to ``on shell'' $n$-point amplitudes.}.

The unconstrained chiral light-cone superfield describes only physical degrees of freedom and in this sense it is as close to the unitarity cut method \cite{Bern:1994zx} as possible. The disadvantage is the absence of manifest Lorentz symmetry, the advantage is the presence of a manifest kinematical supersymmetry, which is realized linearly. Both dynamical supersymmetry and Lorentz symmetry are realized non-linearly.
It is important to keep in mind that the CPT-invariant light-cone superfields are specific for d=4 since they are based on  helicity states for graviton, gravitino, vectors etc. defined  in d=4.

The purpose of this paper is to study the QFT \N=8 SG with  the light-cone superfields. It has been noticed in \cite{Kallosh:2008mq} that the candidate counterterms of \N=8 SG have not been constructed in the light-cone formalism so far. In \cite{Bossard:2009sy} an attempt to convert covariant linearized counterterms into the light-cone ones was made.

In this paper we will continue the investigation of \N=8 SG on the light cone \cite{Kallosh:2008mq}. The formalism which we will develop here does not rely on the existence/non-existence of the counterterms in the light-cone superspace. The main result  will be a prediction  for the outcome of the actual computations using Feynman light-cone  supergraphs.  This result will follow   from the structure of the path integral of the theory and the \N=8 equivalence theorem which will be established here.  The issues  of \N=8 equivalence theorem in the context of anomalies were discussed extensively in \cite{Kallosh:2008mq}. Here we investigate practical consequences of  the light-cone path integral over the unconstrained chiral superfield and \N=8 equivalence theorem, assuming that the theory is anomaly-free.

We will propose here the general formula for the light-cone \N=8 SG generating functional for the amplitudes of any order of perturbation theory and any number of external particles, which comply with translational invariance, linearized supersymmetry
and equivalence theorem.

We will test this functional on the tree and higher loop level and suggest a possibility to use it to investigate the UV properties of the theory. The main observation which will be made in this paper is that the light-cone superfield amplitudes  have a  non-local structure in the transverse directions.
Such a non-local structure is incompatible with the UV divergences. This will lead to  a prediction  of all-loop UV finiteness based on studies of the $n$-point supergraph amplitudes.

{\it The new mechanism which is  the base  of our argument for all-loop finiteness is the increase of the delay of divergences with the increasing number of legs in the loop-amplitudes}. It  has to do with a number of properties of the light-cone supergraphs which we establish in this paper.

a) the dimension of the superfield is zero,  the dimension of the measure in the chiral superspace is zero, therefore the dimension of the light-cone superfield amplitude is zero.

b)  the light-cone superfield is chiral and therefore the superfield  amplitude has to increase the helicity with  each new leg without changing the dimension

c) assuming that the kinematical structure of the loop amplitudes is proportional to the tree amplitudes for each $n$-point amplitude one finds:  the  critical loop order before which the $n$-point amplitude is divergence-free is given by the formula:  $L^{(n)}_{cr}= {n(n-1)+2\over 2}$. It depends strongly on $n$ due to an increasing level of non-locality in transverse directions with the increasing number of legs. This in turn originates from the superfield amplitude properties presented in a) and b). If the assumption of proportionality to tree graphs  is relaxed (see \cite{KR} for details), one finds that $L^{(n)}_{cr}= n+3$. Since it still depends on $n$, the prediction remains valid.

d) at any loop order the divergences should occur at all $n$-point amplitudes with $n\rightarrow \infty$  to preserve the  non-linear \E \, symmetry  of \N=8 SG \cite{Cremmer:1979up,Brink:2008qc,Kallosh:2008ic,Bianchi:2008pu,ArkaniHamed:2008gz,
Kallosh:2008ru}  at a given loop order.

 The delay of divergences increases with the increasing number of legs in the light-cone superfield  amplitude which leads to all-loop finiteness prediction for \N=8 SG.

\section{\N=8 supergravity in the Light-Cone Superspace}

The classical action depends on {\it one unconstrained  scalar superfield}. It can be taken to be either  chiral $\phi$ or antichiral $\bar \phi$. In the real basis, the action depends both on the chiral and anti-chiral superfields \cite{Brink:1982pd},\cite{Brink:2008qc},
\be
S_{cl}^{real} [\phi, \bar \phi(\phi)] ={1\over 2 \kappa^2}  \int d^4x \, d^8\theta \, d^8\bar \theta \,\left(  \bar \phi { \Box \over \partial^{4}_+} \phi -
({ 1 \over \partial^{2}_+} \bar \phi \,  \bar \partial \phi\,  \bar \partial \phi  +cc) +{\cal O}(\phi^4)\right).
\label{action1}\ee
Here the light-cone notation are $ \bar \partial = \partial_1-i\partial_2$, $  \partial = \partial_1+ i\partial_2$, $\partial_{\pm} = \partial_0 \pm \partial_3$. Note that the anti-chiral field is not an independent one, as they both describe  a   single CPT invariant multiplet. The relation between them is given by
$
\bar \phi= {1\over \partial ^{4}_+ } \bar d ^8  \phi
$. Here $\bar d $ are properly defined spinorial covariant derivatives. The action (\ref{action1}) has manifest kinematical supersymmetry associated with eight  $\theta$ and eight  $ \bar \theta$ coordinates of the real light-cone superspace. The other 16 dynamical supersymmetries, Lorentz symmetry and
\E \  symmetry, are realized non-linearly. The total action can be also presented in the form where only the chiral part of kinematical supersymmetry is manifest. This can be achieved by integration over eight $\bar \theta$ coordinates using the relation between chiral and antichiral superfields which allows to express the action in terms of chiral superfields only:
\be
S_{cl}^{chiral} [\phi] ={1\over 2 \kappa^2}  \int d^4x \, d^8 \theta \left( \phi  \Box  \phi -
( (\partial^{2}_+ \phi  ) \bar \partial \phi \, \bar \partial \phi  +...) +{\cal O}(\phi^4)\right).
\label{action2}\ee
The terms with ... also depend only on  chiral superfields $\phi$ but they are more complicated as the cc operation in  (\ref{action1}) gives a dependence on the  anti-chiral superfield. The integration over $\bar \theta$ in this case leads to special kind of differential operators acting on two superfields, see e.g. \cite{Mandelstam:1982cb}, \cite{Belitsky:2004sc} in the YM case. It is a particular combination of space-time $\partial_+$ and spinorial derivatives. In momentum space it is given by a combination $(p^+_{k} \eta_{l}- p^+_{l} \eta_{k})^8$, where $p_k^+\sim {\partial \over \partial x_{+ k}}$ is a momentum of the $k$ particle and $\eta_k \sim {1\over  \sqrt{p^+_k}} {\partial \over  \partial   \theta_k}$ is the super-momentum of the $k$ particle.

In the action (\ref{action2}) only the chiral part of the kinematical supersymmetry is manifest. Thus, when the Feynman supergraph rules are derived on the basis of the action (\ref{action2}), the chiral part of kinematical supersymmetry and $SU(8)$ symmetry are manifest for vertices and propagators of the theory.
For physical observables, like on shell $n$-point amplitudes, one expects that the remaining supersymmetries and the Lorentz symmetry are restored.

The supergraph Feynman rules in the momentum superspace have the following properties: the vertices   will produce some positive powers in momenta $ p_{\bot}= p^1+i p^2$, $\bar p_{\bot}= p^1-i p^2$ and super-momenta $ \eta$. However,    $p^+$ will come in positive as well as negative powers, which leads to non-locality in direction of $x^-$. The non-locality in the non-transverse direction $x^+$ is also possible \cite{Mandelstam:1982cb} due to improved definition of the inverse ${1\over p^+ } \Rightarrow {1\over p^+ +i  \epsilon \; sign \; p^-}$, see Appendix C for details. The superfield   has the  usual massless scalar field propagator proportional to $\Box^{-1}$ as well as spinorial delta-functions.  At present the complete set of the light-cone Feynman supergraph rules was derived in \cite{Brink:1982wv} only for the \N=4 YM case in the real basis based on the action analogous to (\ref{action1}). For our purpose to support the general analysis by sample computations, the set of Feynman rules has to be worked out. Even for \N=4 SYM it will be different from the known ones. The rules have to be in the chiral basis and  in the Fourier superspace. This requires using the external as well as internal lines of the supergraphs characterized by the supermomenta $p_i, \eta_i$ which are a  Fourier transform of the coordinate superspace  $x, \theta$.

The chiral superfield  depends only on physical degrees of freedom:
\bea
&& \phi(x, \theta^a, \bar \theta_a)=e^{{1\over 2} \bar \theta \theta \partial_+} \left ( {1\over \partial ^{ 2}_+} h(x) +\theta^a  {1\over \partial ^{ 3/2}_+}\bar \psi_a(x) + \theta^{ab}  {1\over \partial _{+ }}\bar B_{ab}(x) +\theta^{abc}    {1\over \partial^{ 1/2} _{+ }} \bar \chi_{abc}(x) \right. \nonumber\\
\nonumber\\
&&\left. + \theta^{abcd}  \phi_{abcd} (x) + \tilde \theta_{abc}  \partial^{ 1/2} _{+ } \chi^{abc} + \tilde \theta_{ab} \partial _{+} B^{ab}(x) + \tilde \theta_a \partial^{ 3/2}  _{+} \psi^a(x) + 4 \tilde \theta \partial ^{2}_+ \bar h(x)\right ) \ .
\label{phi}\eea
In the chiral basis it
depends only on  8 chiral Grassmann variables $\theta^a, a=1,...,8$, all dependence on $\bar \theta$ can be absorbed in the change of the basis
\bea
 \phi(x, \theta) =e^{-{1\over 2} \bar \theta \theta \partial_+}\,\phi(x, \theta^a, \bar \theta_a)\ .
\label{phi1}\eea
Here
$
\theta^{a_1...a_n}\equiv {1\over n!}\, \theta^{a_1}... \theta^{a_2}$ and $ \tilde \theta_{a_1...a_n}\equiv \epsilon_ {a_1...a_n b_1... b_{n-8}}\, \theta^{b^1... b^{n-8}}$.
Here $h, \bar h$ are the two helicity states of the graviton, $\bar \psi_a,  \psi^a$ are the two helicity states of the gravitino,
$\bar B_{ab},  B^{ab}$ are the two helicity states of the graviphoton, $\bar \chi_{abc}, \chi^{abc}$ the two helicity states of the graviphotino, and $ \phi_{abcd} $ is a  scalar field. {\it The propagating chiral superfield $ \phi(x, \theta)$ is totally unconstrained}.

Since all gauge symmetries are fixed in the light-cone superspace, the path integral for a chiral scalar superfield is simple.\footnote{In comparison,  gauge symmetries in component $x$-space \N=8 supergravity  \cite{Cremmer:1979up} include  general covariance, local Lorentz symmetry, local supersymmetry and local $SU(8)$-symmetry. This requires sophisticated quantization procedure, which makes Feynman rules rather complicated. Indeed,  even in \N=1 supergravity, the $x$-space Lorentz covariant Feynman rules are quite complicated, see e.g. \cite{Kallosh:1978de}.} We integrate over the unconstrained massless chiral scalar superfield:
\be
Z[J]= e^{i W[J]}= \int d\phi~  e^{i (S_{cl}[ \phi(x, \theta)] +\int d^4 x d^8 \theta  J(x, \theta) \phi(x, \theta))}.
\ee
When $\int J\phi $ is replaced by $\int \phi_{in}(x, \theta) \vec{ {\Box}} \phi$ we have a generating functional for the S-matrix on-shell amplitudes:
\be
\left( e^{i W[\phi_{in}]}\right)_{{\cal N}=8} = \int d\phi~  e^{i (S_{cl}[ \phi(x, \theta)] +\int d^4 x d^8 \theta  \phi_{in}(x, \theta) {{\Box}} \phi(x, \theta))}.
\label{pathint}\ee
 The functional
$W[\phi_{in}]$ describes all connected amplitudes when expanded in powers of $\phi_{in}$. The relevant S-matrix elements include all contributions to the $n$-point amplitudes, 1P irreducible as well as 1P reducible ones.

It is convenient to use the Fourier transform\footnote{Related equations for \N=4 SYM theory are given in Appendix A.} for the light-cone superfield   (\ref{phi1}), (\ref{phi})  multiplied on $\partial^{-2}_+$
\be
 \Phi (p, \eta) = \int d^4 x d^8 \theta~ e^{-ipx -\eta_a (p^{+})^{1/2} \theta^a}  \partial ^{ -2} _+ \phi(x, \theta) \ .
\label{F}\ee
The light-cone superfield $ \Phi (p, \eta) $ has now  the following simple form:
\bea
&& \Phi (p, \eta) = \bar h(p) + \eta_a \psi ^a (p) + \eta_{ab} B^{ab}(p) + \eta_{abc} \chi^{abc}(p)+ \nonumber\\
\nonumber\\
 &&+ \eta_{abcd} \phi^{abcd} (p)+ \tilde\eta^{abc} \bar \chi_{abc}(p) + \tilde \eta^{ab}\bar B_{ab}(p) + \tilde \eta^{a}\bar \psi_{a}(p) + \tilde \eta h(p) \ .
\label{Phi}\eea
Here
$
\eta_{a_1...a_n}\equiv {1\over n!}\, \eta_{a_1}... \eta_{a_2}$ and $ \tilde \eta^{a_1...a_n}\equiv \epsilon^ {a_1...a_n b_1... b_{n-8}}\, \eta_{b^1... b^{n-8}},
$.
This superfield has helicity  +2 and mass dimension -4. When $p^2=0$, it is an on shell superfield in terms of which we will define the generating amplitudes for the S-matrix. On the other hand, when $p^2\neq 0$, it is an off-shell unconstrained superfield, the integration variable in the path integral.
This means that in the momentum space the unconstrained superfield  $ \Phi(p, \eta)$ with $p^2\neq 0$ depends on the momentum $p_{\alpha \dot \beta} \neq \lambda_\alpha \bar \lambda_{\dot \beta}$, and the momentum $p_\mu$ cannot be represented by the pair of commuting spinors.
The free chiral superfield satisfies the equation  $\Box \phi_{in}(x, \theta) =\partial_\mu\partial^\mu \phi_{in}=0$. In momentum space $(p^+ p^- - p_{\bot} \,  {\bar p_{\bot}}) \Phi_{in}=0$. Here $p^\pm $ are $p^0\pm p^3$ and
$p_{\bot}$ and $\bar p_{\bot}$ are  $p^1\pm i p^2$ components of momenta. This free superfield in the momentum space $ \phi_{in} (p, \eta)$ can be represented as a function of commuting spinors,  $p_{\alpha \dot \beta} = \lambda_\alpha \bar \lambda_{\dot \beta}$.

\subsection{Relation to Nair's Superfield and Supertwistors}

It has been noticed by Nair \cite{Nair:1988bq} that the superfield in his construction is very similar to the light-cone superfield of Mandelstam \cite{Mandelstam:1982cb}. In the previous section we  found an exact relation between the light-cone superfield in \N=8 SG and Nair's  type \N=8  superfield, see eq. (\ref{F}).  In the next subsection we will find the relevant generating functions for the amplitudes. In \N=4 SYM one has to multiply the light-cone chiral superfield of
\cite{Brink:1982pd}, \cite{Mandelstam:1982cb}, \cite{Brink:1982wv} by $\partial^{-1}_+$ and perform the Fourier transform in 4 chiral $\theta$'s analogous to (\ref{Phi}) to get eq. (7) in \cite{Nair:1988bq} .  (For   \N=4 SYM, the relevant equations are given in Appendix A.)  We will further discuss the relation between the Nair generating functional and the light-cone path integral in Sec. 3.

One can also relate our light-cone superfield $\Phi (p, \eta)$ in  (\ref{Phi}) to the supertwistor space superfield in eq. (92) of \cite{Mason:2008jy} (and in eq. (212) of \cite{Mason:2009sa}) for the self-dual \N=8 SG. It is given by the following expression:
\be
{\cal H}(Z, \psi) =h(Z) + \psi^A \lambda_A +...+ (\psi^8) \tilde h(Z)\ .
\ee
Up to a change in notation, it is clearly a structure analogous to our eq. (\ref{Phi}). It is also important that this superfield is holomorphic in anti-commuting variables  $\psi$ as ours is holomorphic in $\eta$ and depends on 8 of them. The Chern-Simons type supertwistor action for \N=8 self-dual supergravity  has a simple compact form \cite{Mason:2007ct}, whereas the light-cone superspace action (\ref{action2}) is rather complicated. However, it produces all-loop Feynman supergraphs in Minkowski space signature. Therefore the light-cone superfield path integral is relevant for computations of all quantum corrections to classical \N=8 SG amplitudes and will lead to UV predictions.

\subsection{Light-Cone Generating Functional}
We propose to use the following form of the  generating functional in the chiral light-cone Fourier superspace $W[ \Phi_{in}] = \sum_{n=1}^{\infty} W^n[\Phi_{in}]$
\be \boxed{
 W^n= \prod_{i=1}^{n}  \left (\int d^4p_i \delta (p_i^2) d^8 \eta_i \,  \Phi _{in}(p_i, \eta_i)\right ) \, \delta^4 \left(\sum_{k=1}^{k=n} p_k \right) ~ \delta^8 \left (\sum _{l=1} ^{l=n} (p_l ^{+})^{1/2} \eta_l\right ) {\cal A}_n^{lc}
 (p_i;  \eta_i)}
\label{func}\ee
The origin of the $\delta^4$ and $\delta^8$ functions in (\ref{func}) is from the integration over the coordinate superspace, $x$ and $\theta$,
\be
\delta^4 \left(\sum_{k=1}^{k=n} p_k \right)~ \delta^8 \left (\sum _{l=1} ^{l=n} (p_l ^{+})^{1/2} \eta_l\right )\sim \int d^4 x d^8 \theta ~e^{\sum_k( -i p_k x -\eta_{ak} (p_k^{+})^{1/2} \theta^a)} \ .
\label{12}\ee
Here the contribution from each superfield is taken into account following the single superfield Fourier transform formula (\ref{F}). Since the superfield satisfies the equation $ p_i^2 \Phi _{in}(p_i, \eta_i)=0$, the integration over each momenta is restricted to $d^4 p \; \delta(p^2)$.
The chiral light-cone superspace amplitude is given by the following expression:
\be  \label{A}\boxed{
{\cal A}_n^{lc}
 (p_1,...p_n;  \eta_1, ..., \eta_n)= \left ( \sum _{l=1} ^{l=n} { p_{\bot l}\over (p_l ^{+})^{1/2}} \eta_l\right )^8\,  {\cal P} (p_1,...p_n;  \eta_1, ..., \eta_n)}
\ee
{\it Since the superfield is a scalar without any indices,  the amplitude in eq.  (\ref{func})  is totally symmetric under the permutation of any pair $i\leftrightarrow j$}. It is a polynomial in $8n$ Grassmann variables  $\eta_{i a}$ of maximal degree $8n-32$.
\be
 {\cal P} (p_1,...p_n;  \eta_1, ..., \eta_n)=  {\cal P}^0 + {\cal P}^8+... + {\cal P}^{8n-32} \ .
\label{polyn}\ee
Here ${\cal P}^{0}$ describes the  MHV (maximum helicity violating) amplitude,  ${\cal P}^{8}$ describes the NMHV (next to MHV) amplitude, ${\cal P}^{16}$ describes the NNMHV(next to next to MHV), etc. Each new term in this expression is manifestly $SU(8)$ invariant and therefore comes with 8 extra $\eta$'s.

The functional $W[\Phi_{in}]$ has to have zero dimension and zero chirality/helicity. The measure of integration times the superfield at each point  $\int d^4p_i d^8 \eta_i \,  \Phi (p_i, \eta_i) $ has  dimension 0 and  helicity  -2 at each point. The term  $\delta^4 (\sum_{k=1}^{k=n} p_k )\  \delta^8 \left (\sum _{l=1} ^{l=n} (p_l ^{+})^{1/2} \eta_l\right )$ has zero dimension and helicity $+4$ as a reflection of the chiral superspace measure $\int d^4 x d^8 \theta$. Finally, the amplitude ${\cal A}_n$ is dimensionless, has helicity -4 and +2 at each point. It is a product of 2 terms, the first term $\left ( \sum _{l=1} ^{l=n} { p_{\bot l} \over (p_l ^{+})^{1/2}} \eta_l\right )^8$ has helicity -4 and dimension +4.

{\it This leaves the second term, ${\cal P}(p_1,...p_n;  \eta_1, ..., \eta_n)$,  with the negative total mass dimension -4 and helicity +2 at each point}. This is a very strong restriction on the kinematics of possible amplitudes. Increasing helicity  can be accomplished either by  multiplication on a square bracket or by division on an angular bracket. The first one, multiplication on a square bracket, does not produce non-localities, however, it increases the dimension. Therefore to avoid changing the dimension one has to use both: multiplication on a square bracket and  division on an angular bracket. It is the division on the angular bracket which entails a proliferation of the transverse momenta in the denominator. This in turn leads to proliferation of non-localities with the increasing number of legs and will become one of the important ingredients for the UV analysis of the light-cone superfield amplitudes.

The meaning of the momenta $p_i$ and super-momenta $(p_i ^{+})^{1/2}\eta_i$  in the amplitude ${\cal A}_n^{lc}
 (p_i;  \eta_i)$, where $i=1,...,n$,
 is the following. One considers the Fourier transform to coordinate space for each particle, namely one replaces the superfield $\Phi (p_i, \eta_i)$ by an integral over the coordinate space superfield  $\phi(x_i, \theta_i)$,
 \be
\Phi (p_i, \eta_i) \Rightarrow \int d^4 x_i d^8 \theta_i ~e^{-ip_ix_i -\eta_{ia} (p_i^{+})^{1/2} \theta^a_i}\  \partial ^{ -2} _{+i} \phi(x_i, \theta_i) \ .
\label{Fi}\ee
In this expression one can realize  $p_i$ as $\sim { \partial \over  \partial x_i}$ and $(p_i ^{+})^{1/2}\eta_i$ as $\sim { \partial \over  \partial \theta_i}$.  If the amplitude ${\cal A}_n^{lc}
 (p_i;  \eta_i)$ depends on positive powers of   $p_i$ and  $\eta_i$  (it may  depend on  negative powers of $p_i^+$),  the Fourier transform to coordinate space leads to a local function of superfields except in the direction $x^-$ (and $x^+$, see Appendix B). We can  use the 12-dimensional delta-function as shown in eq. (\ref{12}) and perform an integration over all momenta $p_i$ and super-momenta $\eta_i$.
 The answer will be  a single integral over $x$ and $\theta$ depending on superfields  $ \phi(x, \theta)$, where each superfield  has been differentiated over its coordinates $x, \theta$ in agreement with the polynomials in $p_i, \eta_i$ in the amplitude ${\cal A}_n^{lc}
 (p_1,...p_n;  \eta_1, ..., \eta_n)$.
In a symbolic form, we get
\be
S_n \sim \int d^4 x d^8 \theta ~ D_1\phi(x, \theta) D_2 \phi(x, \theta)...D_n \phi(x, \theta) \ ,
\label{Sn}\ee
where $D_i$ are some polynomials in transverse directions $ { \partial \over  \partial x},  { \partial \over  \partial \bar x}$  and  spinorial directions $ { \partial \over  \partial \theta}$.  Some non-locality in $x^{\pm}$ directions is possible, e.g.   terms like $\epsilon (x^{\pm})$ are possible. For MHV amplitudes the total power of $\theta$-derivatives is 8 since it comes from the term $\left ( { p_{\bot l} \over (p_l ^{+})^{1/2} } \eta_l \right )^8 $, for NMHV is will be 16, NNMHV will have 24, etc.

 If the amplitude ${\cal A}_n^{lc}
 (p_i;  \eta_i)$ depends on negative powers of  $ p_{\bot i}, \bar p_{\bot i} $, the Fourier transform will not lead to a local in $x, \theta$-space expression of the kind presented in eq. (\ref{Sn}).  According to Feynman supergraph rules, the UV divergent loop integrals  can only be of the form  (\ref{Sn}).  Therefore the presence of negative powers of transverse momenta in the light-cone amplitudes indicates that they are not UV divergent.

Our proposal for the light-cone generating functional in \N=8 SG is inspired by the corresponding constructions in \N=4 YM theory \cite{Nair:1988bq,Witten:2003nn,Drummond:2008vq} and its development for the covariant amplitudes in \N=8 SG in \cite{Bianchi:2008pu,ArkaniHamed:2008gz,Kallosh:2008ru}. The major difference with all existing constructions is that it is given by the standard QFT path integral (\ref{pathint}) in terms of a classical action depending on the unconstrained chiral light-cone scalar superfield. Computing quantum corrections means integrating over the chiral light-cone scalar superfield.

\subsection{Supersymmetry and Translational Symmetry}

Under 32 linearized supersymmetries the light-cone superfield transforms as follows
\be
\delta \Phi(p, \eta)= (\epsilon^{\alpha a} q_{a\alpha} + \bar \epsilon^ {\dot \alpha}_a \bar q_{\dot \alpha}^a )\Phi(p, \eta)
=(\epsilon^a \eta_a + \bar \epsilon_{a}  {\partial \over \partial \eta_a} )\Phi(p, \eta)\ .
\ee
Here
\be
q_{a\alpha} = \lambda_\alpha \eta_a\, , \qquad  \bar q^a_{\dot \alpha}= \bar \lambda_{\dot \alpha}{\partial \over \partial \eta_a}\, , \qquad \epsilon^a\equiv  \epsilon^{\alpha a} \lambda_\alpha \, , \qquad \bar \epsilon _a \equiv \bar \epsilon^{\dot \alpha}_a \bar \lambda _{\dot \alpha} \ .
\label{susy}\ee
Here the commuting spinors $\lambda_\alpha,  \bar \lambda_{\dot \alpha}$ in the light-cone superspace depend on $p^+, p, \bar p$, where $p^2= p^+ p^- - p\bar p=0$. In particular,
\be
\lambda_\alpha =  {1\over \sqrt{p^+}} \left( \begin{array}{c}  p_{\bot} \\
 p^+ \\ \end{array} \right)\, , \qquad \bar \lambda_{\dot \alpha } =  {1\over \sqrt{p^+}} \left( \begin{array}{c} {\bar  p_{\bot}} \\
p^+ \\ \end{array} \right)\,.
\label{lambda}\ee
In the light-cone formulation of  \cite{Brink:1982pd},
 \cite{Brink:2008qc} the kinematical supersymmetry is $q_{a 2}$ and $  \bar q_{\dot 2}^a$ and the dynamical is $q_{a 1}$ and $  \bar q_{\dot 1}^a$.
The  functional (\ref{func}), (\ref{A}) is invariant under translation and all 32 linearized supersymmetries. We can present it in the form
\be
 W^n[\Phi_{in}]= \prod_{i=1}^{n}  \left (\int d^4p_i \delta(p_i^2)d^8 \eta_i \,  \Phi (p_i, \eta_i)\right ) \delta^4\left ( P_{\alpha \dot \alpha} \right ) \delta^{16}(Q_{a \alpha}) \,
 {\cal P} (p_1,...p_n;  \eta_1, ..., \eta_n) \ .
\label{func1}\ee
It is manifestly invariant under translation and half of supersymmetry due to conservation of the momenta and super-momenta:
\be
P_{\alpha \dot \alpha}\equiv \sum_{i=1} ^{n} \lambda_{i\alpha}  \bar \lambda_{i\dot \alpha} \  , \qquad Q_{a\alpha}\equiv \sum_{i=1}^{n} \lambda_{i \alpha}  \eta_{i a} \  .
\ee
Indeed, one can easily see that
$
P_{\alpha \dot \alpha} \delta^4\left ( P_{\alpha \dot \alpha} \right )= Q_{a\alpha}   \delta^{16}(Q_{a \alpha})=0
$.
The second part of supersymmetry,
\be
\bar Q_{\dot \alpha}^a \equiv \sum_{i=1}^n \bar \lambda_{i \dot \alpha}{\partial \over \partial \eta_{ai}} \  ,
\ee
is manifest only for the MHV part,
since the action of $\bar Q_{\dot \alpha}^a$ on $\delta^{16}(Q^a_\alpha)$ is proportional to $P_{\alpha \dot \alpha}$ and therefore
\be
\bar Q_{\dot \alpha}^a  \delta^4\left ( P_{\alpha \dot \alpha} \right ) \delta^{16}(Q_{a \alpha}) \,
 {\cal P} (p_1,...p_n;  \eta_1, ..., \eta_n) =0 \ .
\ee
The second part of supersymmetry $\bar Q_{\dot \alpha}^a$ requires non-trivial restrictions on the non-MHV amplitudes which are polynomial in $\eta$ in eq. (\ref{func}):
\be
\bar Q_{\dot \alpha}^a \, \left(  \delta^4\left ( P_{\alpha \dot \alpha} \right ) \delta^{16}(Q_{a \alpha}) \,
\sum_{m=1}^{n-4} {\cal P}^{8m} (p_1,...p_n;  \eta_1, ..., \eta_n)\right) =0 \ .
\label{second}\ee
When $ \bar \lambda_{\dot \alpha}{\partial \over \partial \eta_a}$ acts on the $\eta$-dependent terms in ${\cal P}^8+... + {\cal P}^{8n-32}$,  these polynomials in $\eta$ have to satisfy the condition (\ref{second}). Here ${\cal P}^{8}$ is the NMHV amplitude, ${\cal P}^{16}$ is the NNMHV, etc.

\section{\N=8 Equivalence Theorem}

The form of the light-cone superfield generating functional presented in eq. (\ref{func}),  (\ref{A}) has a remarkable property. On one hand it can be presented by Fourier transform in terms of the on-shell light-cone superfield  $\phi $, \cite{Brink:1982pd},
 \cite{Brink:2008qc} given in  eq. (\ref{phi}). It also has an  explicit non-covariant  $p^+$ and $p_{\bot}$ momenta in the spinorial $\delta$-functions, as shown in eq. (\ref{func}).  However, the Fourier transform of the superfield  $ \Phi= (p^+)^{-2}  \phi$  in Fourier space depends only on physical states of external particles as shown in (\ref{Phi}).
Moreover, all explicit Lorentz non-covariant factors in the  spinorial $\delta$-functions in  (\ref{func}),
 (\ref{A}) can be presented in a Lorentz covariant form since
\be
\delta^8 \left (\sum _l (p_l ^{+})^{1/2} \eta_l\right ) \left ( \sum _k  { p_{\bot l} \over (p_k^{+})^{1/2}} \eta_k\right )^8 =\prod_{a=1}^{a=8} \sum_{k>l\geq 1}^{n}\langle k l \rangle \eta_{ka} \eta_{la}= \delta^{16} (\sum _i  \lambda^\alpha_i \eta_{ai})  , \quad \alpha=1,2.
\ee
Here the first Lorentz non-covariant 8-dimensional $\delta$-function originates from the Fourier transform from $\theta$-space. The second Lorentz non-covariant 8-dimensional $\delta$-function is a factor in the amplitude  (\ref{A}) responsible for the restoration of the dynamical supersymmetry and Lorentz symmetry. We may rewrite these two $\delta$-functions as follows
\be
\delta^8 \left (\sum_l \lambda^1_l  \eta_l \right )  \delta^8 \left (\sum _k \lambda^2_k  \eta_k\right ) =\left(\sum_l  \sum_k \lambda^1_l  \lambda^2_k \eta_{la}   \eta_{ka}\right)^8 \ ,
\ee
where $\lambda^1_i= (p_l ^{+})^{1/2}$ and $\lambda^2_i= { p_{\bot l} \over (p_l ^{+})^{1/2}}$ are
commuting spinors $\lambda, \bar \lambda$ as shown in eq. (\ref{lambda}). Note that when a particular $\eta_{la}$ from the first $\delta^8$ is multiplying $\eta_{ka}$ from the second $\delta^8$, the product $ \eta_{la}   \eta_{ka}$ with the same $SU(8)$ index $a$  becomes antisymmetric in $l,k$: $ {1\over 2} (\eta_{la}   \eta_{ka}-\eta_{ka}   \eta_{la})$. This picks up the angular brackets in front of such products, $\langle lk\rangle$, and the 16-dimensional  $\delta$ becomes Lorentz covariant.
Thus we have used the fact that the product of  two eight-dimensional delta functions, one  in  (\ref{func}) and  another one in (\ref{A}), can be brought to the form depending only on angular brackets where the antisymmetric product in the light-cone gauges is
\be
\langle k l \rangle \equiv { p_{\bot k} p_l^+ -  p_{\bot l} p^+_k\over (p_k^+ p_l^+)^{1/2}}=  \lambda_{\alpha i} \lambda^\alpha_j \ .
\label{angular}\ee
This means that  the light-cone generating functional   (\ref{func}) can be brought to the Lorentz-covariant  form
\be
 W^n[\Phi_{in}]= \prod_{i=1}^{n}  \left (\int d^4p_i \delta(p_i^2) d^8 \eta_i \,  \Phi (p_i, \eta_i)\right ) \, \delta^4\left (\sum_{m=1}^{m=n} p_m\right ) \prod_{a=1}^{a=8} \sum_{k>l\geq 1}^{n}\langle k l \rangle \eta_{ka} \eta_{la}\,
 {\cal P} (p_1,...p_n;  \eta_1, ..., \eta_n) \ .
\label{func2}\ee

We see that in  the final form of the functional in eq. (\ref{func2}) all possible miracles took place: the superfield which we are using, shown in eq. (\ref{Phi}), has no explicit dependence on non-covariant momenta despite the fact that it has a simple relation to the light-cone superfield in  coordinate space, which looks highly non-Lorentz covariant, see eq. (\ref{phi}). The product of kinematic supersymmetry $\delta^8$-function depending on $p^+$ in eq. (\ref{func}) and dynamic supersymmetry $\delta^8$-function depending on $p_{\bot}$   shown in eq. (\ref{A}) depends only on the angular spinorial brackets, $\langle ij\rangle$ defined in the light-cone gauges as shown in eq.
(\ref{angular}).  However, as soon as the answer depends on the angular brackets  $\langle ij\rangle$, there is no need to specify it for the light-cone expression (\ref{angular}), it is a Lorentz covariant object.

The next important ingredient of our formula is the following: the light-cone amplitude  (\ref{A}),
\be {\cal A}_n^{lc}
=  \left(\sum_l  \lambda^2_l \eta_l \right)^8\ {\cal P}^{lc} (p_1,...p_n;  \eta_1, ..., \eta_n)\ ,
\label{A1}\ee
consists of the Lorentz non-covariant factor $ (\sum_l  \lambda^2_l \eta_l )^8$ and the factor  $ {\cal P}^{lc} (p_1,...p_n;  \eta_1, ..., \eta_n)$. This function is a polynomial in $\eta$, totally symmetric in $n$ particles. The first term presents the MHV ``all-plus helicity amplitude,'' the last one -  the anti-MHV
``all-minus helicity amplitude''.

Since we have shown that the generating functional (\ref{func2}) has all other factors Lorentz covariant, it means that
the value of the factor in the light-cone amplitude $ {\cal P} ^{lc}(p_i;  \eta_i)$  must be exactly  equal to  the corresponding totally symmetric in $n$-particle expression  ${\cal P}^{cov}
 (p_i;  \eta_i)$  in covariant gauges.  Thus, the  condition for the gauge-independence of the on-shell amplitudes between the light-cone and covariant gauges requires that
\be \boxed{
{\cal P}^{lc} (p_1,...p_n;  \eta_1, ..., \eta_n)={\cal P}^{cov}
 (p_1,...p_n;  \eta_1, ..., \eta_n)}
\ee

One may use the equivalence theorem as follows: whenever we have information on a covariant part of the ``all-plus helicity amplitude''  $ {\cal P} (p_1,...p_n;  \eta_1, ..., \eta_n)$, we can insert it into our functional in eqs. (\ref{func}),  (\ref{A}).
This is a result which has to come out when the computations are performed using the light-cone supergraph technique.
It is interesting here that the newly established amplitude in the light-cone formalism has +2 helicity at each point and is totally symmetric  in  all $n$ points due to the fact that the \N=8 SG light-cone superfield is a scalar without any indices. In the context of the covariant generating functional described in \cite{Bianchi:2008pu, Kallosh:2008ru}, this ``all-plus helicity amplitude''  was defined as ${\cal P}(+  + ...+ +)$ in \cite{Kallosh:2008ru}, following the \N=4 YM construction in  \cite{Drummond:2008vq}.

The meaning of this function was not clear in the previous work: there is no gravity amplitude with all-plus helicity. Only when this function  ${\cal P}(+  + + +)$ was multiplied by a factor $\langle 12\rangle^8$ it became a (- - + +...  ++ ) gravity MHV amplitude, which has a physical meaning.  Here we have shown that this totally symmetric function ${\cal P}(+  + ...+  +)$  has a very simple physical meaning: it is a factor in the amplitude (\ref{A}), (\ref{A1}) in the light-cone superfields  of \N=8 supergravity. The total symmetry follows from the scalar property of the light-cone superfield as well as from the symmetry of the other factor $ (\sum_l  \lambda^2_l \eta_l )^8$ in the amplitude under interchange of any two particles.

Our equivalence theorem also explains the relation between the Nair's covariant generating functional for \N=8 SG described in \cite{Bianchi:2008pu, Kallosh:2008ru} for the MHV amplitudes and the regular QFT supergraph part integral in the light-cone superspace given in (\ref{pathint}),(\ref{func2}). In both cases, one can use the knowledge of the gravity amplitude, e.g. $ M^n(- - + +...+ +)$, and relate it to the ``all-plus helicity amplitude''  ${\cal P}(+  + + +)$:
\be
M^n(- - + +...+ +) = \langle 12\rangle^8 {\cal P}(+  + + +)\ ,
\ee
and vice versa. All other $n$-point amplitudes in the Nair's type procedure for \N=8 in \cite{Bianchi:2008pu, Kallosh:2008ru} for particles other than gravitons are obtained by a certain fermionic differentiation. In the light-cone formalism one has to integrate over $\eta$ in eqs. (\ref{func}), (\ref{func2}) and use the expansion of the superfield in component fields given in eq. (\ref{Phi}): this is a regular superfield procedure which follows from the manifest supersymmetry.

\section{4-Point Light-Cone Amplitude}

The 4-point light-cone superfield amplitude has only the MHV part,
\be {\cal A}_4^{lc}
=   (\sum_{l=1}^4  \lambda^2_l \eta_l )^8 {\cal P}^{0} (++++) \ .
\label{A4}\ee
Here $ {\cal P}^{0}$ is the `` all-plus helicity amplitude'' $ {\cal P}^{0} (++++)$ as defined in \cite{Kallosh:2008ru}. For the tree approximation
\be
{\cal P}^{0} _{tree}(++++)=i   {1\over \kappa^2 stu}{[ij ]^4\over \langle i'j'\rangle^4 } \ .
\ee
Here  $i'\neq i,  i'\neq j$ and $j'\neq i, j'\neq j$. It has mass dimension -4 and helicity +2 at each point, as required. Note that ${[ij ]^4\over \langle i'j'\rangle ^4}$
 is equal to any combination out of 4 non-coinciding points since
\be
{[34 ]\over \langle 12\rangle } ={[24 ]\over \langle 13\rangle }={[14 ]\over \langle 32\rangle } ={[23 ]\over \langle 14\rangle }={[13 ]\over \langle 42\rangle } ={[12 ]\over \langle 43\rangle}
\label{ID}\ee
due to momentum conservation.
In the  1-loop approximation \cite{Bern:1998ug,Naculich:2008ew,Kallosh:2008ru}
\be
{\cal P}^{0} _{1loop}(++++)=i {[ij ]^4\over \langle i'j'\rangle^4 } \left({\cal I}_4^{(1)} (s, t) + {\cal I}_4^{(1)} (s,u)+ {\cal I}_4^{(1)} (t,u)\right) \ .
\ee
Here ${\cal I}_4^1 (s, t)$ is the so-called box diagram. It can be represented in the following form
\be
{\cal P}^{0} _{1loop}(++++)=i {[ij ]^4\over \langle i'j'\rangle^4 }\left( { f\over tu} + {g\over us} +  {h\over st}\right) \ .
\ee
Here $f, g, h$ are some dimensionless functions of Mandelstam variables and $1/\epsilon$ parameter which is  taking care of the IR property of the box diagram. Here again,  ${\cal P}^{0} _{1loop}(++++)$ has mass dimension -4 and helicity +2 at each point, as expected.

In the 2-loop approximation there are double box diagrams and non-planar diagrams:
\be
{\cal P}^{0} _{2loop}(++++)=i \kappa^2 {[ij ]^4\over \langle i'j'\rangle^4 } \left( s^2  ({\cal I}_4^{(2)} (s, t) + {\cal I}_4^{(2)} (s,u)) + ...\right) \ ,
\ee
where ... means a cyclic permutation of $s, t, u$. When the corresponding 2-loop graphs are computed one finds \cite{Bern:1998ug,Naculich:2008ew}
\be
{\cal P}^{0} _{2loop}(++++)=i \kappa^2 {[ij ]^4\over \langle i'j'\rangle^4 }\left( {f'\over t}  + {g'\over s} +  {h'\over u}\right) \ ,
\ee
where $f', g', h'$ are some dimensionless functions of Mandelstam variables and $1/\epsilon$ parameter.  Here   ${\cal P}^{0} _{2loop}(++++)$ has mass dimension -4 and helicity +2 at each point, as it should.

\subsection{3-Loop Case}

In the 3-loop approximation there are triple box diagrams and non-planar diagrams:
\be
{\cal P}^{0} _{3loop}(++++)=i \kappa^4 {[ij ]^4\over \langle i'j'\rangle^4 } X(s,t,u) \ .
\ee
The dimensionless function $X(s,t,u)$ is given by two groups of terms \cite{Bern:2007hh}. One group is proportional to the 4-th power of Mandelstam variables, some combination of $s_{ij}^4$,  times a non-local integral of the triple box type, as well as non-planar graphs. This integral is well convergent in UV, it has mass dimension -8 and upon computation has to produce a term of the form $ s_{kl}^{-4}$ up to dimensionless combinations of $s/t$ etc. The second group is proportional to the third power of Mandelstam variables, some combination of $s_{ij}^3$,  times a non-local integral of the triple box type, as well as non-planar graphs. This integral is also well convergent in UV, it has mass dimension -6 and upon computation has to produce a term of the form $ s_{kl}^{-3}$ up to dimensionless combinations of $s/t$ etc. In both cases the  integral is non-local. If one tries to relate it to possible UV divergences by expanding in a power series in momenta, one finds that there are following terms
\be
{s_{ij}^3\over \Lambda^6} \ , \qquad  {s_{ij}^4\over \Lambda^8} \ ,\qquad {s_{ij}^5\over \Lambda^{10}} \ ,\qquad etc. \ ,
\ee
whereas the terms
\be
s_{ij}^0 \ln \Lambda \ , \qquad  {s_{ij}\over \Lambda^2} \ ,\qquad {s_{ij}^2\over \Lambda^{4}}
\label{super} \ee
are missing. This absence of  UV divergence $\ln \Lambda$ means finiteness at the 3-loop level. In addition, the computations established the absence of other two terms in eq. (\ref{super}) , ${s_{ij}\over \Lambda^2}$ and $ {s_{ij}^2\over \Lambda^4}$, which   is called superfiniteness.  From the perspective of the light-cone formalism one may interpret the result of the 3-loop computation as follows: the function $X(s,t,u)$ has dimension zero, however, it is a non-local function of momenta given by the ratio's of the type ${s_{ij}^3\over s_{kl}^{3}}$ times the dimensionless functions of $s/t$ etc. In the momentum Fourier space $x$ it cannot be presented by a local function in transverse directions, but it is well described in terms of the light-cone superfields with some non-local amplitude.
\be
({\cal A}_4^{lc})^{3loop}
\sim    \left(\sum_{l=1}^4  \lambda^2_l \eta_l \right)^8 \kappa^4 {[ij ]^4\over \langle i'j'\rangle^4 } X(s,t,u)
\label{A4a}
\ee
A well known 3-loop UV divergence for the 4-graviton amplitude can be reproduced upon $\eta$-integration  of the underlying light-cone superfield amplitude under condition that
\be
X(s,t,u)\sim \ln \Lambda
\ee
However, the presence of such terms  would contradict the light-cone supergraphs where the superfield amplitude would be
\be \boxed{({\cal A}_4^{lc})^{3loop}_{UV}
\sim \left(\sum_{l=1}^4  \lambda^2_l \eta_l \right)^8  \kappa^4 {[ij ]^4\over \langle i'j'\rangle^4 } \ln \Lambda }
\label{A43loop}
\ee
Under Fourier transform to $x, \theta$ space, the expression in   (\ref{A43loop}) as a function of light-cone superfields is non-local, it cannot appear as a local divergence in the supergraphs since it is not expressible as a function of the light cone-superfields (and their derivatives in the transverse space-time and spinorial directions) at one point in $x, \theta$ space.

To make it more clear, we can rewrite the $\ln \Lambda $ part of the amplitude in a form in which its singularity in the momentum space is given in terms of the Mandelstam variables. Here we have to take into account that ${\cal P}^{0} _{3loop}(++++)=i \kappa^4 {[ij ]^4\over \langle i'j'\rangle^4 } X(s,t,u)$  has dimension -4 and helicity + 2 at each point. We find that the 3-loop UV divergence, which could be a candidate for the UV divergent 4-point amplitude, corresponds to the light-cone superfield amplitude $ {\cal A}_4^{lc}
=   (\sum_{l=1}^4  \lambda^2_l \eta_l )^8 {\cal P}^{0} (++++)
$, where
\bea
{\cal P}^{UV}_{3loop}  (1^+, 2^+, 3^+, 4^+)\sim
 \kappa^4  \left ( {[12 ] [ 34]  \over s }+ {[13 ] [ 24] \over t } +{[14 ] [ 23] \over u } \right) {N^*(4)\over stu} \ln \Lambda \ .
\label{sing}\eea
Here $N(4)= \prod_{i=1}^{i=3} \prod_{j=i+1  }^{4} \langle j k\rangle $ and $N^*(4)= \prod_{i=1}^{i=3} \prod_{j=i+1  }^{4} [ j k] $ were introduced in
\cite{Berends:1988zp}.
However, the supergraphs will not produce $\ln \Lambda$ divergence with the non-polynomial dependence on Mandelstam variables in the amplitude. This is in agreement with the 3-loop computation \cite{Bern:2007hh} where such terms are absent.

The square of the Bel-Robinson tensor $(R_{\alpha\beta\gamma\delta} \bar R_{\dot \alpha\dot \beta\dot \gamma\dot \delta})^2$
 has a long history in studies on SG counterterms. It was proposed as a candidate 3-loop counterterm for \N=1 SG in \cite{Deser:1977nt}. It was also generalized to the case of \N=8 SG in \cite{Kallosh:1980fi}, \cite{Howe:1981xy}. Meanwhile we have learned that the local counterterm candidate, a supersymmetric version of the  square of the Bel-Robinson tensor failed and did not show up as a logarithmic divergence in the 3-loop computations in \cite{Bern:2007hh}. Here we will show that the light-cone supergraphs predicted this.

As we explained above the light-cone  candidate of the 3-loop divergence in  the Fourier momentum space is given by
\be
 W^{3loop}_{lc}[\Phi_{in}]_{UV}\sim  \prod_{i=1}^{4}  \left (\int d^4p_i \delta(p_i^2)d^8 \eta_i \,  \Phi (p_i, \eta_i)\right ) \, \delta^4\left (\sum_{m=1}^{m=n} p_m\right ) \prod_{a=1}^{a=8} \sum_{k>l\geq 1}^{n}\langle k l \rangle \eta_{ka} \eta_{la}\,
\kappa^4  {[ij ]^4\over \langle i'j'\rangle^4 }\ln \Lambda \ .
\label{3loop}\ee
Here we have combined both spinorial $\delta^8$-functions into a $\delta^{16}\sim  \prod_{a=1}^{a=8} \sum_{k>l\geq 1}^{n}\langle k l \rangle \eta_{ka} \eta_{la}$ which is useful for the evaluation of the amplitude.

Now we have to check that
  its gravitational part is related to the linearized square of the Bel-Robinson tensor (with the on-shell graviton).
It has been explained in \cite{Kallosh:2008mq} that  in helicity formalism  the square of the Bel-Robinson tensor is given by the following expression:
\be
\kappa^{4} \int d^4 x (R_{\alpha\beta\gamma\delta} \bar R_{\dot \alpha\dot \beta\dot \gamma\dot \delta})^2   \Rightarrow \kappa^4 \prod_{i=1}^{4}  (\int d^4p_i  ) \;  \langle 12\rangle^4 [34 ]^4 \bar h(p_1)  \bar h(p_2)  h(p_3)  h(p_4) \ .
\ee
Let us find this expression now as a part of the 3-loop light-cone ``counterterm'' (\ref{3loop}).
\begin{enumerate}

\item  From the $\delta^{16}$  function $\prod_{a=1}^{a=8} \sum_{k>l\geq 1}^{n}\langle k l \rangle \eta_{ka} \eta_{la} $
we pick up the term with only $\eta_1^8$ and $\eta_2^8$.

\item We take ${\cal P}^{0} (1^+, 2^+, 3^+, 4^+)$ in one of the equivalent forms, ${[34 ]^4\over \langle 12\rangle^4 }$.

\item Perform the $\eta$ integration $\prod_{i=1}^{4} \int  d^8 \eta_i $
\end{enumerate}
\be
 W^{3loop}_{lc}[\Phi_{in}]\sim  \prod_{i=1}^{4}  \left (\int d^4p_i \delta(p_i^2) d^8 \eta_i \,  \Phi (p_i, \eta_i)\right ) \, \delta^4\left (\sum_{m=1}^{m=4} p_m\right ) \langle 12 \rangle^8 \eta_{1}^8 \eta_{2}^8\,
 \kappa^2   {[34 ]^4\over \langle 12\rangle^4 } +...
\label{part}\ee
Keeping in mind that
\be
\Phi (p, \eta) \sim  \bar h (p)+....+ \eta^8  h(p) \ ,
 \ee
 we have to take $h(p_1)$ and $h(p_2) $ from the first 2 superfields, but $\bar h(p_3)$ and $\bar h(p_4) $ from the other two to provide all required $4\times 8$ powers of $\eta$. This gives us
\be
 W^{3loop}_{lc} [h_{in}, \bar h_{in}]\sim  \prod_{i=1}^{4}  \left (\int d^4p_i \delta(p_i^2) \right) \, \delta^4\left (\sum_{m=1}^{m=4} p_m\right ) (h(p_1) h(p_2) \bar h(p_3) \bar h(p_4) )
 \kappa^4  \langle 12\rangle^4   [34 ]^4  +...
\ee
The terms ... above include all supersymmetry partners of the square of the Bel-Robinson tensor. Therefore our expression (\ref{3loop})  would be a correct candidate for the 3-loop counterterm in the light-cone superspace.
However in the light-cone superspace the amplitude  (\ref{A43loop}) contains $ \kappa^4  {[ij ]^4\over \langle i'j'\rangle^4 }$. This expression, or its analog  shown in eq.  in (\ref{sing}), is non-local in transverse space-time directions.

\subsection{Local Versus Non-Local and UV Divergences}

The supergraph rules in the chiral $p, \theta$ space  will provide positive powers of $p_{\bot}, \bar p_{\bot}$ and $ \eta \sim {\partial\over \partial \theta}$. The propagators will produce the $\theta$ $\delta$-functions and terms $1/l^2$ where $l$ are momenta carried by the propagators of the internal lines.    They  correspond to the combinations of integration variables $q$ and momenta of external particles $p_1, p_2$, like $l=q+ p_1+p_2$.  The integral over loop momenta $q_1, ... ,q_L$ will have positive powers of internal and external transverse momenta and propagators in the denominator. Such integrals cannot produce terms
$
 {1\over s^2tu} \ln \Lambda$ which are required to support the $\ln$ divergence of the 3-loop 4-point amplitude.
For example, the 3-loop amplitudes of \N=8 SG have the form \cite{Bern:2009kf}:
\be
\prod_{L=1}^{L=3} d^4 q_L \; {N(q_L, p_i)\over \prod_{m=1}^{m=10} l_m^2} \ .
\label{Zvi}\ee
Here $q_1, q_2, q_3$ are three independent loop momenta and  $N(q_L, p_i)$ are polynomials in loop momenta and external momenta.\footnote{In the light-cone supergraphs the $\pm$ directions do not have to be non-polynomial since the vertices have terms like ${1\over p^+ +i\epsilon \; sign \;  p^-}$.}  When the propagators $l^{-2}=(q+ p_1+p_2)^{-2}$ are expanded like
\be
{1\over (q+ p_1+p_2)^2}= {1\over q^2} \left (1+{2q(p_1+p_2) + 2p_1p_2\over q^2}\right) ^{-1}\approx {1\over q^2} \left (1-{2q(p_1+p_2) + 2p_1p_2\over q^2}+...\right)
\label{exp}\ee
the first term $\sim {1\over q^2}$ may lead to a UV divergence but the second term has an extra factor ${1\over q}\sim  {2q(p_1+p_2) + 2p_1p_2)\over q^2}$ and the degree of divergence is reduced comparative to the first one by one power of $\Lambda$. Higher orders of expansion will have higher powers of external momenta ${p\over q}$ and eventually will become finite due to increasing number of powers of the integration variable $q$ in the denominator.

 In the final answers for the amplitudes after integration one can  have terms  containing factors like 1/s, 1/t, 1/u    as well as terms like $\ln s/t$. However, one cannot have 1/s, 1/t, 1/u  in the expression for the loop integrals from which we extract the
 counterterms/logarithmic divergences. There is standard all-loop renormalization procedure in QFT based on Bogoliubov-Parasiuk-Hepp-Zimermann theorem which states that in general QFT to any order in perturbation theory all divergences are removed by the counterterm vertices corresponding to superficially divergent integrals. The procedure involves an expansion of the INTEGRAND of the loop integral like the one in eqs. (\ref{Zvi}), (\ref{exp}):  one finds terms which are divergent, and the next terms in the expansion of the integrand are finite. It  boils down to the following: the UV divergences (the counterterms) must be locally constructed, i.e. written without using such operators as $1/\Box $. The only exception in the light-cone gauges is the presence of factors like ${1\over p^+ +i\epsilon \; sign \; p^-}$, see Appendix C for details.

 Thus the divergence (\ref{A43loop}), (\ref{sing}) should not appear in the Feynman supergraph computations of the 3-loop order in \N=8 SG. According to equivalence theorem, the same should happen in a covariant computation, as it was indeed discovered in \cite{Bern:2007hh} via the unitarity method.

 There is one more puzzle on the square of the Bel-Robinson tensor which is now resolved by the light-cone supergraphs.  It was shown in  \cite{Kallosh:2007ym} that the non-local 1-loop effective 4-point amplitude is given by the linear in gravitons part of the following
 expression
\be
S^{\rm box}=\int {d^4x_1 d^4x_2  d^4x_3  d^4x_4 \over x_{12}^2 x_{23}^2  x_{34}^2  x_{41}^2 } \left(R_{\alpha \beta \gamma \delta}  (x_1)  R_{\dot \alpha \dot \beta \dot \gamma \dot  \delta}  (x_2) R^{\alpha \beta \gamma \delta}  (x_3)  R^{\dot \alpha \dot \beta \dot \gamma \dot  \delta}  (x_4) + sym.\right)
\label{BR}\ee
 The \N=8 supersymmetric version of the 1-loop on-shell effective action (\ref{BR}) is rather simple \cite{Kallosh:2007ym}:
\be
 S^{\rm box}_{{\cal N}=8} \sim \int d^4 x_1 d^4 x_2  d^4x_3  d^4 x_4 \; d^{16} \theta_B \;  {W_1 W_2  W_3 W_4  \over x_{12}^2 x_{23}^2  x_{34}^2  x_{41}^2 } \ ,
\label{8} \ee
where $W_a\equiv W(x_a,\theta_B), a=1,2,3,4$. Here the index $B$ means either a specific choice of the basis in $\theta$-space  found in \cite{Kallosh:1980fi} or a representation $[232,848]$ of the $SU(8)$ used in \cite{Howe:1981xy}. Eq.  (\ref{8})  provides  an \N=8 supersymmetric extension to the 1-loop box diagram for the square of the Bell-Robinson term with each curvature spinor $R$ at the different corner of the box, as one can see in eq. (\ref{BR}).  Thus the non-local amplitudes do realize the \N=8 supersymmetrization of the non-local square of the Bell-Robinson term.

The local expression for the 3-loop counterterm is
 \be
S_{{\cal N}=8}^{3loop}= \kappa^{4} \int d^4 x\ d^{16}  \theta_{B}\   W^4(x, \theta_B)  = \kappa^{4} \int d^4  x   R_{\alpha \beta \gamma \delta} (x)  R_{\dot \alpha \dot \beta \dot \gamma \dot  \delta}  (x) R^{\alpha \beta \gamma \delta}  (x)  R^{\dot \alpha \dot \beta \dot \gamma \dot  \delta}  (x)+... \ .
\ee
 The local expression for square of the Bell-Robinson term is forbidden by the light-cone supergraph rules. However, the non-local amplitude is not forbidden by the structure of the light-cone amplitudes and therefore we find a complete consistency in this picture. The \N=8 supersymmetric version of the non-local square of the Bell-Robinson term remains a relevant important structure of the \N=8 SG amplitudes.

 There is an interesting connection of these structures with the projective superspace and supertwistor expressions developed in \cite{Hatsuda:2008pm}.

 \subsection{Changing Dimensions}

In the unitarity cut method \cite{Bern:1994zx,Bern:1998ug}
the continuation to arbitrary dimensions can be done by analytic continuation of the bosonic integrals for the 4-point graviton amplitudes.
For example the 3-loop computation in \N=8 SG \cite{Bern:2007hh} provides an integral in dimension  $d$ of the form
\be
\int d^d q_1 d^d q_2 d^d q_3 {q_i^2 s_{ij}^3\over l_i^2... (l+p)^2} \ ,
\label{d}\ee
where there are 10 propagators in the denominator. For $d=4, $ the UV behavior is ${s_{ij}^3\over \Lambda^6}$ +..., where the terms ... have higher powers of momenta and inverse powers of $\Lambda$.
For $d=6$, the integral is $\sim s_{ij}^3 \ln \Lambda$. Therefore the conclusion in   \cite{Bern:2007hh}  is that \N=8 SG in $d=6$ is divergent at 3-loops.

The question is: Can we use the light-cone formalism developed here for d=4 to predict the 3-loop logarithmic divergence in d=6 or 1-loop divergence in d=8?

The kinematical part of the 4-graviton amplitude in d=4 is given in the helicity formalism as
$\langle 12\rangle^4 [34]^4$.  In the coordinate space this corresponds to the square of the Bel-Robinson tensor, $(R_{\alpha\beta\gamma\delta} \bar R_{\dot \alpha\dot \beta\dot \gamma\dot \delta})^2$,  where $R_{\alpha\beta\gamma\delta}$ and $ \bar R_{\dot \alpha\dot \beta\dot \gamma\dot \delta}$ are curvature spinors of opposite chirality. In higher dimensions one can represent the square of the Bel-Robinson tensor  as a particular fourth  power of the curvature tensor, $t_8 t_8 R^4$,  which avoids the chiral notations of d=4 theory.
\be
d=4, \quad (R_{\alpha\beta\gamma\delta} \bar R_{\dot \alpha\dot \beta\dot \gamma\dot \delta})^2 \quad \Rightarrow \quad t_8^{\mu_1...\mu_8} t_8^{\nu_1...\nu_8} R_{\mu_1\mu_2\nu_1\nu_2}R_{\mu_3\mu_4\nu_3\nu_4} R_{\mu_5\mu_6\nu_5\nu_6} R_{\mu_7\mu_8\nu_7\nu_8}, \quad d\neq 4 \ .
\ee

This fact was used in computations,  starting from \cite{Bern:1998ug},  of \N=8 SG amplitudes in various dimensions between 4 and 11. It was sufficient to analytically continue the range of integration from 4 to d in the integrals like (\ref{d}). The possibility of such an analytic  continuation in dimensions when using  the unitarity cut method in maximally supersymmetric theories is supported by the fact that the  states of d=11, \N=1 supergravity are identical in all dimensions, with  only the names of the states changing as  dimension is changed.

For our purpose of extending our light-cone d=4 supergraph analysis to d=6 or d=8 we have to find a 6-dimensional analog not of the 4-graviton amplitude $\sim (R_{\alpha\beta\gamma\delta} \bar R_{\dot \alpha\dot \beta\dot \gamma\dot \delta})^2$ but of the light-cone superfield amplitude:
\be
  \kappa^4 \left (\sum_{l=1}^4  {p_{\bot  l}\over \sqrt {p^+_l }} \eta_l \right )^8  {  \langle i'j'\rangle^4 [ij ]^4\over \langle i'j'\rangle^8 } \ .
  \ee
  The familiar term here, which is easy to continue to other dimensions, is $\langle i'j'\rangle^4 [ij ]^4$. This can be replaced by  $t_8 t_8 R^4$ in other dimensions. This still leaves us with  problems.  The first one is the chiral delta function $\left (\sum_{l=1}^4  {p_{\bot  l}\over \sqrt {p^+_l }} \eta_l \right )^8$, which depends on transverse  momentum $p_{\bot  l}= (p_1+i p_2)_l$ at each point $l$. In d=6 there are 4 transverse momenta $p_1, p_2, p_4, p_5$, in d=8 there are 6 transverse momenta, so the delta function in the light-cone amplitude is problematic. Another problem is the term ${  1\over \langle i'j'\rangle^8 }$. In d=4 it can be presented as shown in eq. ({\ref{angular}) and it also depends on $p_{\bot } $ at the points $i'$ and $j'$. It is therefore not clear if such an analytic continuation of the d=4  light-cone amplitude can be defined for d=6, 8 and what the answer is.

  Therefore the best we know at present is that our light-cone superfield amplitude in d=4  is not obviously generalizable to any dimension $d\neq 4$. To test our argument about the non-locality of the light-cone amplitude versus available computations first we have to construct the relevant formalism in d=6, 8 and/or find an analytic continuation of the light-cone amplitude.
 Before this is done properly,  the light-cone superfield predictions of d=4 are not useful for $d\neq 4$.

\subsection{4-Point Amplitude at  Higher Loops}

The 4-point $(- - + +)$ L-loop graviton amplitude has to be proportional to the tree level 4-point amplitude times the $s, t, u$ polynomials as explained in \cite{Bern:1998ug}.
It also follows from the fact that the non-local square of the Bel-Robinson tensor  can be  made \N=8 supersymmetric using the covariant linearized superfield as shown in example (\ref{8}). Insertions of $s, t, u$ polynomials into eq.  (\ref{8}) will preserve the linearized supersymmetry.

No other structures of this kind are available. We can therefore multiply the tree level amplitude by a dimensionless factor $\kappa^{14} stu (s^4+t^4+u^4)$ which, as we will explain, is the first polynomial which will remove the non-locality factors from the light-cone superfield 4-point amplitude. It corresponds to the  7-loop case since we will have $\kappa^{14}$ times $\kappa^{-2}$ at the tree amplitude which results in $\kappa^{12}= \kappa^{2(L-1)}$.
\be
{\cal P}^{0} _{7loop}(++++)\sim \kappa^{14} stu (s^4+t^4+u^4) {\cal P}^{0} _{tree}(++++)\sim
  \kappa^{12} (s^4+t^4+u^4) {[ij ]^4\over \langle i'j'\rangle^4 } \ .
\ee

Although it looks like this amplitude is still non-polynomial in momenta, one can in fact rewrite it as follows:
\be
{\cal P}^{0} _{7loop}(++++)\sim
  \kappa^{12}\left( s^4 {[34 ]^4\over \langle 12\rangle^4 } +t^4 {[13 ]^4\over \langle 24\rangle^4 }+ {[14 ]^4\over \langle 23\rangle^4 }\right) .
\ee
Here we used relations (\ref{ID}). Now in each term we can reorganize things using the fact that $s_{ij}^4= \langle i j \rangle^4 [ij]^4$. We find a polynomial in momenta expression which becomes a candidate counterterm for the 7-loop  4-point amplitude:
\be
{\cal P}^{0} _{7loop}(++++)\sim
  \kappa^{12}\left( {[34 ]^4[ 12]^4 } + {[13 ]^4 [ 24]^4 }+ {[14 ]^4 [ 23]^4 }\right) \ .
\ee
This expression is totally symmetric in exchange of any pair  of particles, it has mass dimension -4 and helicity +2 at each point.
It will correspond to a covariant counterterm $\kappa^{12} \int d^4 x d^{16}\theta_B W^4 \partial^{8}$ where $\partial^{8}$ is a symbolic expression for a symmetric combination of derivatives acting on each of the four superfields $W(x, \theta_B)$ \cite{Kallosh:1980fi}. The gravitational part in the momentum space is the square of the Bel-Robinson tensor with insertions of 4 powers of symmetric Mandelstam variables.

\section{$n$-Point Loop Amplitudes, Proportional to Tree Ones}

The set of tree level amplitudes in \N=8 SG has a non-linear \E \, symmetry which relates amplitudes with different number of points. Therefore,   {\it as a simplest probe} of the n-point UV divergences we may assume that, as in the 4-point case, the candidate counterterms are some polynomials in particle momenta times the tree diagrams
(later we will relax this assumption).

For the $n$-point light-cone amplitudes  in eq. (\ref{polyn}) ${\cal P} (p_1,...p_n;  \eta_1, ..., \eta_n)$ we have to consider the polynomial in $\eta$'s, \,
  ${\cal P}^0 + {\cal P}^8+... + {\cal P}^{8n-32}$. The first term, the MHV amplitudes, is the only one for $n=4$. For higher $n$ we have to study the properties of all terms in eq. (\ref{polyn}). We may start with the MHV tree amplitude derived in \cite{Berends:1988zp},
  \be
  {\cal P}^{0}_{tree} (1^+, 2^+, 3^+, 4^+,....,n^+)= {[12] [n-2\, n-1]\over \langle 1 n-1\rangle N(n)} \times f (p) \ .
 \label{BGK} \ee
Here $N(n)= \prod_{i=1}^{i=n-1} \prod_{j=i+1  }^{n} \langle j k\rangle $ and $f(p)$ is some polynomial function of momenta and angular brackets. To make this amplitude non-singular, we can multiply it by $[ 1 n-1]\rangle N^*(n)$. In such case we find
  \be
  {\cal P}^{0}_{tree} (1^+, 2^+, 3^+, 4^+,....,n^+)= {[12] [n-2\, n-1]  [ 1 n-1] N^*(n)\over \kappa^2 s_{1, n-1}  NN^*(n)}\times f (p) \ .
  \ee
  The singularity is now given in terms of invariants $s_{ij}$. The power of momenta in $s_{1, n-1}  NN^*(n)$ is equal to $n(n-1)+2$. In case of $n=4$ we would have $t^3 u^2 s^2$, which gives a power of momenta equal to $3\times 2+2 =14$. After symmetrization, the power may be higher, however, at this point the only important fact is that the power grows as $n^2$ at large $n$ for the amplitude $ {\cal P}^{0}_{tree} (1^+, 2^+, 3^+, 4^+,....,n^+)$.

  For NMHV case  $ {\cal P}^8$ and NNMHV case  $ {\cal P}^{16}$ etc. a separate study it required. There is an increasing amount of information on these amplitudes,  \cite{Bianchi:2008pu,ArkaniHamed:2008gz,Drummond:2009ge,Mason:2009sa,
ArkaniHamed:2009si},  which hopefully will help to evaluate the light-cone UV divergences in \N=8 SG.

We may also use the recently discovered supersymmetric recursion relations \cite{ArkaniHamed:2008gz} valid for the tree amplitudes. We can  relate them to the class of loop amplitudes which have the same helicity structure and infer the situation with the proliferation of the singularities with growing number of legs in the amplitude.  The recursion
 relations are presented in a  form useful for us in eq. (24) of   \cite{ArkaniHamed:2008gz}. Basically, they state that the $n$-point tree amplitude may be generated using the previously known bosonic BCFW deformation
\cite{Britto:2005fq}, $\lambda _1(z)=\lambda_1+z \lambda_2$ and $\bar \lambda _2(z)=\bar \lambda_2-z \lambda_1$, as well as a fermionic one, $\eta_1(z) = \eta_1+z \eta_2$. The corresponding recursion relations are
\bea
&&M(\{\eta_1(z), \lambda_1(z), \bar \lambda_1\}, \{ \eta_2, \lambda_2, \bar \lambda_2(z)\}, \eta_i)
\nonumber\\&&=\sum_{L,R} \int d^8 \;  \eta M_L (\{\eta_1(z_P), \lambda_1(z_P), \bar \lambda_1\}, \eta, \eta_L) {1\over P^2(z)} M_R (\{\eta_2, \lambda_2, \bar \lambda_2(z_P) \}, \eta, \eta_R) \ .
\eea
An important information for us in these recursion relations has to do with the following fact. The 4-point amplitude can be given via a sum of various 3-point amplitudes connected by one line between them, see e.g. Fig. 1 in \cite{ArkaniHamed:2008gz}. To add one leg one can add one more 3-point vertex connected to the 4-point one by a propagator. On can add a leg in many different ways. For a general case one can view the left part of the graph in Fig. 1 in \cite{ArkaniHamed:2008gz}  as consisting from various 3-point amplitudes connected by propagators, the same for the right-hand side. In all cases, adding one more leg reduces to the following: pick up any of the existing $n-1$ legs in the amplitude, let us call it leg $i$,  and create a 3-point amplitude  with two external legs, one is a new one, $n$, and the other one is the one we called $i$. The third leg of the 3-point amplitude is the propagator, connecting $n$ and  $i$ legs to the remaining $n-2$ legs. This will increase a singularity of the $n$-point amplitude comparative to $(n-1)$-point amplitude by one factor of $(p_n+p_i)^{-2}$. However, there are $n-1$ such possibilities since $i=1, 2..., n-1$. This means that a symmetrized $n$-point amplitude has an increased
number of singularities in momenta, comparative to the $(n-1)$-point amplitude: the $n$-point amplitude has an extra factor
comparative to the $(n-1)$-point amplitude:
$
M_n \sim {M_{n-1}\over (s_{ij})^{n-1}}
$.
Let us compare this proliferation of singularities with the one we deduced in the previous subsection from the explicit Berends-Giele-Kuijf  formula (\ref{BGK}) for the MHV amplitudes. We found that the power of momenta in the denominator in $n$-point amplitude  is equal to $[n(n-1)+2]$. This means that the proliferation of the singularities between  the $n$-point amplitude and   the $n-1$-point amplitude is
$
[n(n-1)-2]-[(n-1)(n-2)-2]= 2(n-1)
$.
This is in full agreement with the prediction of the recursion relation. However, recursion relations are not constrained by the MHV amplitude, they are generic.
We conclude that the $n$-point MHV and non-MHV amplitudes, proportional to tree ones, may start having  divergences from the loop $L_{cr} = {n(n-1)\over 2} +1$.

\section{General Helicity Structures in $n$-Point Amplitudes}

In loop amplitudes some new helicity structures are possible, which are not necessarily proportional to tree amplitudes. Consider a case when at some loop order $L$ a certain $m$-point amplitude can be given in the form of a local expression, without non-local factors, and therefore may serve as a candidate for a divergence.

Now we may try to consider at the same loop order the amplitude with $m+n$ external legs. Since we are not changing the loop level, we have to keep the dimension of the amplitude without change, however, we have to increase the helicity of the amplitude by a factor of +2 at each of the $n$ new legs.

 The only way to increase helicity without changing dimension is to multiply by factors like $[ij]$ and divide by exactly the same number of factors $\langle i'j' \rangle$. It is therefore impossible to  avoid  a dependence on transverse directions in the denominator of the amplitudes with additional legs. Therefore, even if we have a candidate divergence at the loop $L$ with $m$-point amplitude, the $m+n$ amplitude at the loop $L$ cannot be divergent. It may be divergent at the higher loop level.

  The difference with a simpler case when the loop amplitudes repeat the helicity structure of the tree amplitudes is the following. When $M^{Lloop} \sim M^{tree}f(s_{ij})$ we had explicit information on the properties of the amplitudes, like the delay of divergence to the $L_{cr}={n(n-1)+2\over 2}$. In more general case \footnote{The generic structure of all such local amplitudes will be given in \cite{KR}. } the explicit formula  is $L_{cr}=n+3$, however, we only need the fact explained above that the delay of divergences increases with the number of legs. This is sufficient to proceed with the claim that the delay of the UV divergences due to non-locality of $n$-point amplitudes together with unbroken \E \, symmetry leads to an all-loop finiteness prediction for the \N=8 SG.

\section{\E \, Symmetry and All Loop Finiteness}

Here we would like to put our trust into the validity of the continuous  \E \, symmetry at the perturbative quantum level. At this point we will ignore the puzzle created by the $U$-duality of  string theory and \N=8 QFT and the fact that d=4 black holes in \N=8 supergravity break the continuous classical symmetry $E_{(7,7)}(\mathbb{R})$ to the discrete subgroup $E_{(7,7)}(\mathbb{Z})$ \cite{Kallosh:1996uy}. We will talk about it in the next section.

  At this point one should consider the non-linear symmetries which relate the amplitudes with different number of particles.  If one believes that non-linear symmetry \E \, symmetry is not anomalous, it would require that at any particular loop level, all  $n$-point  amplitudes are divergent. The reason is that   the coset part of the symmetry is non-linear in superfields \cite{Brink:2008qc}. Symbolically,
  \be
  \delta_{E_{7(7)}} \phi(x, \theta) = \sum_{n=0}^{\infty}( f_{abcd}^n   \Sigma^{abcd} +  f^{abcd}_n \bar   \Sigma_{abcd})\phi^n (x, \theta)
\label{coset}  \ee
Here   $f_{abcd}^n$ and $f^{abcd}_n $ depend on $\theta$ and ${\partial\over \partial \theta}$ and $\partial_+$ and $\partial_+^{-1}$. $ \Sigma^{abcd}$ and $ \bar   \Sigma_{abcd}$ are the 70 parameters of the ${E_{7(7)}\over SU(8)}$
 coset transformations. Together with 63 $SU(8)$ transformations they form the 133-parameter \E \, symmetry. However, the 63 $SU(8)$ transformations are realized linearly \cite{Brink:2008qc}
 \be
\delta_{SU(8)} \phi(x, \theta)= \omega^a{}_b T^b{}_a \phi(x, \theta)
 \ee
  where $T^b{}_a$ is some quadratic operation in fermionic differential operators. Thus the diagonal $SU(8)$ part of \E \, is not useful for our purposes, for example, it relates the 4-point amplitudes to the 4-point amplitudes only. However, the coset part of \E \, shown in eq. (\ref{coset}) requires  all $4+m$-point partners with $m\rightarrow \infty$  to satisfy the symmetry requirement.
But according to the discussion of the singularities in the momentum space for   $n$-point amplitudes, this is not possible.

It appears that the current studies of \E \, symmetry in \N=8 d=4 QFT in \cite{Brink:2008qc,Kallosh:2008ic,ArkaniHamed:2008gz,Kallosh:2008mq,Kallosh:2008ru} support the expectation that $E_{(7,7)}(\mathbb{R})$ may be anomaly free in perturbative QFT computations. Therefore by combining the  \E \, symmetry with the supersymmetric recursion relations  we are able to argue that the light-cone supergraphs predict  all-loop finiteness of d=4 \N=8 SG. As we already explained, the non-linear nature of the \E \, symmetry requires that at any loop level we have an $n$-point amplitude partner of the 4-point amplitude. The proliferation of singularities delays the logarithmic UV divergences to higher and higher loops and in the limit $n\rightarrow \infty$ leads to the all-loop finiteness prediction.

\section{Non-Perturbative \N=8 and  Extremal Black Holes}
The assumption/argument  that \N=8 SG is perturbatively finite leads to a number of puzzles with regard to the U-duality of string theory \cite{Hull:1994ys}. Namely, U-duality in the context of string theory means that the $E_{(7,7)}(\mathbb{R})$ symmetry of the classical \N=8 SG is broken down to a discrete subgroup of it, $E_{(7,7)}(\mathbb{Z})$.

Clearly, perturbative \N=8 SG  is not a complete theory. At least one of the reasons is the existence of non-perturbative solutions of \N=8 SG in d=4. Some well known solutions are extremal black holes of \N=8 SG which recently have been related to quantum information theory \cite{Kallosh:1996uy}. Recently some more general non-perturbative solutions of \N=8 SG have been studied  \cite{Bergshoeff:2008be}.

The studies of the QFT amplitudes near the Minkowski space suggest that $E_{(7,7)}(\mathbb{R})$ symmetry of the classical \N=8 SG may be unbroken in the perturbative QFT computations. Meanwhile, the black hole charges are quantized and they form a fundamental representation  56 of $E_{(7,7)}(\mathbb{Z})$.  The symplectic charge matrix-vector  $Q$ for N=8 consists of electric $e_{\Lambda \Sigma}$ and magnetic $m^{\Lambda \Sigma}$ charges forming the fundamental representation of $E_{7(7)}$
\be
Q \equiv (m^{\Lambda \Sigma}, e_{\Lambda \Sigma} ) \ .
\ee
Moreover, near the black hole horizon the values of the scalars are fixed by the attractor mechanism \cite{Ferrara:1995ih} and depend on black hole charges. Quantized charges break $E_{(7,7)}(\mathbb{R})$ symmetry of the background down to $E_{(7,7)}(\mathbb{Z})$ symmetry  and therefore the value of the vev which  the scalar fields  acquire near the black hole horizon is chosen by the background. It corresponds to the minimal value of the \N=8 black hole potential \cite{FK2}:
\be
{\cal V} _{BH}(\phi, Q) = Z_{AB}  Z^{*AB}\ ,  \qquad A,B=1,\dots , 8.
\label{N8pot}\ee
Here $Z_{AB}$ (and its conjugate $ Z^{*AB}$) is the central charge matrix (and its conjugate).
\be
Z_{AB}(\phi, Q) = f_{AB}^{\Lambda \Sigma}e_{\Lambda \Sigma} - h_{\Lambda \Sigma, AB } m^{\Lambda \Sigma}\ ,
\label{CH}\ee
where $Q$ is charge vector, a fundamental $56$ of $E_{7(7)}$, and the bein $f_{AB}^{\Lambda \Sigma}(\phi), \, h_{\Lambda \Sigma, AB } (\phi)$ is an element of the coset space ${E_{7(7)}\over SU(8)}$ connecting
 the  real $56$  to complex $28$ of $[AB]$. It depends on 70 real scalars $\phi^i$, where the local $SU(8)$ symmetry was used to remove 63 scalars from the 133-dimensional representation of  scalars in $E_{7(7)}$.
The derivative of the black hole potential over 70 moduli is given by the following expression
\be
\partial_i {\cal V}_{BH}={1\over 4}  P_{i,[ ABCD]}\Big [Z^{*[CD} Z^{*AB]} + {1\over 4!} \epsilon ^{CDABEFGH} Z_{EF} Z_{GH} \Big ] .
\label{critical1}\ee
Here  $P_{i,[ ABCD]} d\phi^i$ is the  $70\times 70$  vielbein of the ${E_{7(7)}\over SU(8)}$ coset space, $i=1, \dots , 70$,  and it is invertible. This means that the necessary and sufficient condition for the critical points of the black hole potential with regular $70\times 70$-beins is given by an algebraic condition on the central charge matrix:
\be
\partial_i {\cal V}_{BH}^{attr}(\phi, Q) =0 \qquad \Rightarrow \qquad Z^{*[AB}Z^{*CD]}+ {1\over 4!} \epsilon ^{ABCDEFGH} Z_{EF} Z_{GH} =0 \ .
\label{N8}\ee
Various solutions of these equations of the type $\phi^i_{attr} = \phi^i(Q)$, relating the scalar field vev's  to black hole charges are known.

Any of these exact non-perturbative non-linear solutions can be considered as a background in which we compute  quantum corrections. Actually,  all predictions of counterterms which have been made in the background field method have the following property. All divergences can be represented as local expression depending on the background field, e.g. in d=4 pure gravity the 2-loop UV counterterm is \cite{Kallosh:1974yh}
\be
{\kappa^2\over \epsilon (4\pi)^4}{209\over 2880}\int d^4x R_{\mu\nu}{}^{\lambda \delta} R_{\lambda \delta}{}^{ \eta \xi} R_{\eta \xi}{}^{\mu\nu} \ ,
\ee
where $R=R_{\mu\nu}=0$. The coefficient of the divergence, ${209\over 2880}$, does not depend on the choice of the background, it is the same number near the flat background when the 3-point graviton amplitude is computed, or in any  other background which is an  exact solution of classical equations.

The fact that in \N=8 SG the amplitudes near Minkowski space were shown to be UV divergent through 3 loops suggest that the computation of the background functional, for example in the extremal black hole background, will be also free of divergences at least through 3 loops.
The supergraph analysis predicts  that the amplitudes are all-loop finite. This  suggests that the background functional in the classical black hole background may be all-loop finite. The absence of UV divergences (which in our analysis requires the unbroken $E_{(7,7)}(\mathbb{R})$ symmetry) does not seem to be affected by the properties of the black hole background, which breaks $E_{(7,7)}(\mathbb{R})$ down to $E_{(7,7)}(\mathbb{Z})$ due to charge quantization in the background. This is reminiscent of the QFT non-Abelian gauge theories which have the same UV properties independently of the fact that the gauge symmetry may be broken spontaneously \cite{'tHooft:1971rn}.

One of the possible consequences of our results is that one should be able to compute  \N=8 QFT corrections  and use the corresponding effective action to study quantum corrections to classical solutions. For example, the near horizon extremal black hole $AdS_2\times S^2$ geometry has 8 supersymmetries unbroken.  It is not unlikely that  it will be an exact solution of \N=8 SG with the account of quantum corrections. It would be interesting to study this conjecture as well as other properties of perturbative and non-perturbative \N=8 SG.

\section{List of ``Things To Do''}

It would be good to have a better and much more detailed understanding of the \N=8 light-cone superspace and the corresponding d=4 supergraph QFT which would help to confirm the analysis and predictions  of this paper.  In particular:

\begin{itemize}

\item  \N=4 SYM Feynman rules in the momentum superspace were proposed and used for the computations of  dilatation operators \cite{Belitsky:2004sc}. Supergraph light-cone Feynman rules were also used in  \cite{Ananth:2006ac} to compute some diagrams in  an \N=4 SYM theory deformed to \N=1.  The extensive set of light-cone computations was performed in \N=4 SYM theory in components, not manifestly supersymmetric, see \cite{Bassetto:2007zj} and references therein. So far, however, in existing literature there are no explicit supergraph computations of the on shell amplitudes in which we are interested. Such computations would help us to
    see if the generating functional formalism suggested  here for \N=4 SYM and for \N=8 SG is in agreement with actual supergraph computations. Hopefully, the  examples will support the structure of the functional proposed above.

    \item One can start with \N=4 SYM and compute the 4-point supergraph amplitude using the Feynman rules in the chiral superspace. This is expected to produce for the color-ordered light-cone amplitude in any order of perturbation theory of the form         \be
{\cal A}_4^{lc}
 (p_1,...p_4;  \eta_1, ..., \eta_4)_{YM}= \left ( \sum _{l=1} ^{l=4} { p_{\bot l}\over (p_l ^{+})^{1/2}} \eta_l\right )^4  {\cal P} (p_1,...p_4)_{YM} \ .
\label{Ampl4}\ee

Here ${\cal P} (p_1,...p_4)_{YM}= {A(- - + + )\over \langle 12\rangle ^4}+...$, where dots mean symmetrization. Thus any would be local logarithmic divergence in the 4-gluon amplitude $A(- - + + )$ in the underlying light-cone supergraph computation will have an extra $ \langle 12\rangle^ {-4}$. This is a non-local in transverse directions term in supergraphs, which prevents a UV divergence in the 4-gluon amplitude. If one can confirm it for \N=4 SYM, it will indicate that the analogous is likely to be the property of \N=8 SG.

\item In \N=4 SYM the coupling constant in d=4 is dimensionless. Therefore, as opposite to \N=8 SG, the non-locality of the 4-point function cannot be removed by multiplication by a dimensional polynomial of Mandelstam variables times a power of a coupling constant. This means that the non-locality of the 4-point \N=4 SYM light-cone  superfield amplitude makes it finite. In \N=8 SG such dimensionless polynomials are available, one can only delay the infinities to a higher loop level.
    The \E \, symmetry has to work to relate the $n$-point amplitudes to each other. Only in such case the infinities are delayed forever since $n \rightarrow \infty$. This part has to be studied on top of everything we learn from \N=4 SYM.

    \item We argued in Sec. 6  that the helicity structure of the $n$-point on-shell loop amplitudes is the same as the tree amplitudes and the difference comes out only in the dependence on scalar invariants $p_i \cdot p_j=s_{ij}$. We also suggested a more general scenario, that the loop amplitudes are not proportional to tree amplitudes. It would be very useful to check all available computations of the $n$-point loop diagrams both in \N=4 SYM and in \N=8 SG to clarify the situation by specific examples.

\item  The light-cone supergraph formalism was not yet formulated in $d\neq 4$, to the best of our understanding. If the formalism will be constructed one can compare the information on UV divergences in  diverse dimensions. with the light-cone supergraph predictions. In maximal SG there are two most interesting cases. Light-cone supergraphs should allow the d=8 1-loop logarithmic $R^4$ divergence and  in d=6 they should allow the 3-loop logarithmic $\partial^6 R^4$ divergence. If there will be a contradiction,  we will have to revisit  the d=4 predictions and find the source of the problem.

    \item  It requires much more work to establish the absence of anomalies in BRST symmetry, on which the equivalence theorem of Sec. 3 is based. So far we have only partial hints that the theory is anomaly-free.

        \item The considerations in this paper are of a very general nature, not yet supported by specific  computations. Clearly, if we are looking to establish the UV properties of \N=8 SG at the arbitrary loop level, we cannot avoid being very general. However, before the set of arguments in this paper can be fully trusted, one should critically analyze each argument and look for examples/counterexamples via specific computations. This will help to rule out or improve and confirm  each argument one by one.

\end{itemize}

\section{Summary}

In this paper we  identified the structure of the amplitudes  of the  light-cone chiral scalar superfield $\Phi(p_i, \eta_i) $ in \N=8 SG. We have established a relation between the light-cone superfield amplitudes and graviton amplitudes as well as amplitudes of all other particles of \N=8 SG. Upon certain integration over the Grassmann variables these superfield amplitudes reproduce the familiar expressions for the graviton amplitudes. In this sense the light-cone superfields amplitudes  are fundamental as they describe the results of Feynman supergraph computations in unitary gauge of the theory where only  physical degrees of freedom propagate. The graviton and all other particle amplitudes are computable from the superfield amplitudes.

We found that the $n$-point light-cone amplitude in the momentum superspace is given by the following expression:
\be   \boxed{
{\cal A}_n^{lc}
 (p_1,...p_n;  \eta_1, ..., \eta_n)=  \left (\sum _{l=1} ^{l=n} { p_{\bot l}\over (p_l ^{+})^{1/2}} \eta_l\right )^8  {\cal P} (p_1,...p_n;  \eta_1, ..., \eta_n)}
\label{AmplSG}\ee
where
\be
{\cal P} (p_1,...p_n;  \eta_1, ..., \eta_n) = {\cal P}^0 (+ + ....+ +)  + NMHV + NNMHV +... \ .
\ee
Here the symmetric $n$-point expression ${\cal P} (p_1,...p_n;  \eta_1, ..., \eta_n)$ has the total dimension 0 and helicity +2 at each point. The first term ${\cal P}^0 (+ + ....+ +) $ depends only on momenta $p_i$ and represents the MHV amplitudes.  The next terms, from NMHV all the way to  $\overline {MHV}$, are polynomials of increasing power in $\eta$.

These amplitudes turned out to have some remarkable properties.  To reproduce the local counterterms for the graviton amplitudes, the superfield amplitudes have to be non-local in transverse directions, $p_\bot=p_1+ip_2$. The qualitative reason is relatively simple to understand for those familiar with Nair amplitude construction \cite{Nair:1988bq, Witten:2003nn,Bianchi:2008pu,ArkaniHamed:2008gz,Kallosh:2008ru}.

For example,  the light-cone superfield 4-point amplitude $A(p_i, \eta_i) , i=1,2,3,4$ is related to the 4-point MHV 4-graviton amplitude $M^4(- - + +)$ as follows:
\be
 A_4^{lc}(p_i, \eta_i) = \left ( \sum _{l=1} ^{l=4} { p_{\bot l}\over (p_l ^{+})^{1/2}} \eta_l\right )^8  {\cal P} ( + + + +)=  \left ( \sum _{l=1} ^{l=4} { p_{\bot l}\over (p_l ^{+})^{1/2}} \eta_l\right )^8 {M^4(- - + +) \over \langle 12 \rangle^8} \ .
\label{example}\ee
Whenever the graviton amplitude $M^4(- - + +)$ is local (polynomial in transverse momenta $p_\bot$ momenta), the light-cone superfield amplitude may be non-local due to the extra factor ${1 \over \langle 12 \rangle^8}$.  When the denominator in the right hand side of equation (\ref{example}) tends to zero, i.e.
$\langle 12 \rangle =  { p_{\bot 1} p_2^+ -  p_{\bot 2} p^+_1\over (p_1^+ p_2^+)^{1/2}}\rightarrow 0$, the fermionic delta-function $ \left ( \sum _{l=1} ^{l=4} { p_{\bot l}\over (p_l ^{+})^{1/2}} \eta_l\right )^8$ is not vanishing. By  adding to it the well known information about the graviton amplitudes at certain loop order one can predict the UV behavior of the light-cone supergraphs.

The 3-loop divergence of the 4-point amplitude is ruled out since ${\cal P}^0 (+ ++ +)\sim  \kappa^4 {[ij]^4\over \langle i'j' \rangle^4}$. (We explain in Appendix B why in pure gravity the light-cone analysis  does not prevent the 3-loop divergence.) If we study only the 4-point function, ignoring \E \, symmetry, we make the following prediction:  The first polynomial in momenta amplitude with proper helicity and dimension for  the 4-point function is available at the 7-loop level.

Further we argued that the delay of divergences for the 4-point function to higher loop order has to be supported by an analogous delay of the $n$-point functions due to unbroken \E \, symmetry. This symmetry is realized  non-linearly and requires the presence of all $n$-point amplitudes with $n \rightarrow  \infty$. Meanwhile, the structure of the $n$-point   amplitudes suggests a  proliferation of the singularities  in Mandelstam-type  variables   in the $n$-point amplitudes. The qualitative reason for this is that  the light-cone amplitude for increasing number of legs   has an increasing number of helicities +2 at each point and fixed dimension: one cannot increase the number of square brackets $[ij]$ in the nominator, one has to add also angular brackets $\langle kl\rangle $ in the denominator. This in turn means that when  $n$ increases, the delay of divergences  for the $n$-point amplitudes
moves the UV infinities to higher and higher loop order. In the limit $n \rightarrow  \infty$ the number of lops is pushed to infinity and we conclude that  all infinities are delayed for all light-cone superfield loop amplitudes: the light-cone supergraphs predict that  the theory is  UV finite.

\section*{Acknowledgments}

I thank  L. Alvarez-Gaume, L. Brink, D. Freedman, P. Howe,  J. Kaplan,   Sung-Soo Kim,  A. Marrani, B. Nilsson, S. Shenker,  E. Sokatchev,  M. Soroush, P. Ramond, K. Stelle,  S. Stieberger and L. Susskind  for the most useful discussions of \N=8 SG.  I am particularly grateful to
Z. Bern,  L. Dixon, G. Korchemsky, A. Linde  and T. Rube for their help in clarifying the issues raised in this paper. This work is supported by the NSF grant 0756174.

\section{APPENDIX}

\begin{appendix}

\section{\N=4 SYM on the Light Cone}

 We start with the Mandelstam version on \N=4 SYM theory \cite{Mandelstam:1982cb}
given in the useful for us form in \cite{Belitsky:2004sc}. The chiral superfield (with rescaled fermions) of dimension zero and helicity -1 in the chiral basis is
\bea
&&\phi(x, \theta) = \partial_+^{-1} A(x) + \theta^A \partial_+^{-1/2} \bar \psi_A(x) +  \theta^{AB} \bar \phi_{AB} (x)
+\tilde  \theta_A \partial_+^{1/2}\psi^A(x) +\tilde \theta \partial_+ \bar A(x) \ .
\eea
 The \N=4 SYM action is
\begin{align}
  S=& - {1\over g^2} \int d^4 x d^4 \theta \left ({1\over 2} \phi^a \Box \phi^a +{2\over 3} f^{abc} \partial_+ \phi^a \bar \partial \phi^b \phi^c \right.+ {2\over 3} [ \partial_+^{-2}\phi^b, \partial  \partial_+^{-2} \phi^c]\cr
 & \left.-{1\over 2} f^{abc} f^{ade} \left \{ \partial_+^{-2} (\phi^b \partial_+ \phi^c )[ \partial_+^{-2}\phi^d, \partial  \partial_+^{-1} \phi^e] -{1\over 2} \phi^b \phi^d [ \partial_+^{-2}\phi^c, \partial  \partial_+^{-2} \phi^e]  \right \} \right) \ ,
\end{align}
where $[\phi_1, \phi_2] \equiv \prod_{A=1}^4 \left (\partial_{\theta^A}^{(1)} \partial_+^{(2)}-\partial_{\theta^A}^{(2)} \partial_+^{(1)}\right ) \phi_1 \phi_2$ and $\prod_{A=1}^4 \partial_{\theta^A}=  \partial_{\theta^1}...  \partial_{\theta^4}$. Note that in N=4 SYM $A=1,2,3,4$ is an $SU(4)$ index and $a$ is the non-Abelian color index.

Now we introduce the Fourier of the rescaled superfield by
\be
 \Phi (p, \eta) = \int d^4 x d^4 \theta ~e^{-ipx -\eta_a (p^{+})^{1/2} \theta^a}  \partial ^{ -1} _+ \phi(x, \theta) \ ,
\label{FYM}\ee
The light-cone superfield $ \Phi (p, \eta) $ has now  the following simple form:
\bea
 \Phi (p, \eta) = \bar A(p) + \eta_a \psi ^A (p) + \eta_{AB} \phi ^{AB}(p) + \tilde \eta^{A} \bar \psi_A (p)+  \tilde \eta A(p) \ .
\label{PhiYM}\eea

The generating functional of the light-cone superfield amplitudes can be expressed in term of the superfield (\ref{PhiYM}) and the rest of the discussion of \N=8 SG can be easily applied to \N=4 SYM.

The main result of this paper is easy to formulate using, for simplicity, the analogous structure in \N=4 YM theory, which is more familiar than \N=8 SG.  We start with the eq. (2.42) of Witten's paper \cite{Witten:2003nn}, which presents a useful form of Nair's construction \cite{Nair:1988bq} for MHV  amplitudes:
\be
\hat A \sim \delta^4(P) \delta^8(\Theta)  {\cal P}^0 (\lambda_i) \ ,
\ee
where
\be
 \delta^4(P)= \int d^4x ~\exp({ix\sum_i p_i} ) \ , \quad
\delta^8(\Theta)= \int d^8 \theta^a_\alpha~ \exp ({i\theta^a_\alpha \sum_i \eta_{ia} \lambda^\alpha_i}) \ , \quad a=1, ..., 4 \ , \quad \alpha=1,2.
\ee
Here $  {\cal P}^0 (\lambda_i)= {1\over {\cal N}(123...n)}$ for the tree approximation where ${\cal N}(123...n)\equiv \langle 12\rangle .... \langle n1 \rangle$. It has been explained in \cite{Drummond:2008vq} that including the non-MHV  amplitudes requires to replace ${\cal P}^0 (\lambda_i)$ by a sum of terms which are polynomial in $\eta^4$, namely, the complete tree level amplitude is
\be
\hat A =i g^{n-2}(2\pi)^4 \delta^4(P) \delta^8(\Theta)  \left({\cal P}^0 (\lambda_i) +{\cal P}^4 (\lambda_i, \eta_i)+...+{\cal P}^{4n-16}  (\lambda_i, \eta_i)\right) \ ,
\label{N4}\ee
where ${\cal P}^{4m}  (\lambda_i, \eta_i)$ has $4m$ powers of $\eta$.

Our observation translated into \N=4 YM theory is the following. One can derive eq. (\ref{N4})  from the QFT path integral where the integration variable is the unconstrained chiral light-cone superfield of \N=4 YM theory.
\be
 \left(e^{i W[\phi_{in}]}\right)_{{\cal N}=4}= \int d\phi~  e^{i (S_{cl}[ \phi(x, \theta)] +\int d^4 x d^4 \theta  \phi_{in}(x, \theta) {{\Box}} \phi(x, \theta))} \ .
\label{pathintYM}\ee
The action depends only on a {\it chiral light-cone superfield} $\phi(x, \theta)$ depending on 4 variables $\theta$, see the details in the Appendix A. Therefore the Fourier transform has to be used only for 4 $\theta$'s, not for 8.  We have to split the Lorentz covariant $\delta^8(\Theta)$ into two parts
\be
\delta^8(\Theta)=\left( \int d^4 \theta^a_1 \exp( i\theta^a_1 \sum_i \eta_{ia}  \lambda^1_i) \right)  \times \left( \sum_i \eta_{ia}  \lambda^2_i\right)^4 \ ,
\ee
where $\lambda^1=\sqrt {p^+}$ and $\lambda^2= {p_{\bot}\over \sqrt {p^+}}$. The first $\delta^4$ is used for the Fourier transform whereas the second one is a part of the amplitude. In the Fourier space the amplitude associated with the path integral (\ref{pathintYM}) is
\be \boxed{
{\cal A}_n^{lc}
 (p_1,...p_n;  \eta_1, ..., \eta_n)_{YM}= \left ( \sum _{l=1} ^{l=n} { p_{\bot l}\over (p_l ^{+})^{1/2}} \eta_l\right )^4  {\cal P} (p_1,...p_n;  \eta_1, ..., \eta_n)_{YM} } \ .
\label{Ampl}\ee
Here ${\cal P} (p_1,...p_n;  \eta_1, ..., \eta_n)_{YM}$ has dimension -1 and helicity +1 at each point to compensate the \N=4 SYM light-cone superfield $\Phi(p_i, \eta_i)$ at each point in the effective action.

The light-cone amplitude (\ref{Ampl}) at the level of loop corrections  may, in principle, have terms with only positive powers of $p_{\bot}$, $\eta$ (inverse powers of $p^+$ are possible in the light-cone supergraphs). Such terms may appear in Feynman supergraph computations as UV $\ln \Lambda$ divergences. However, a direct inspection of the amplitude (\ref{Ampl}) shows that it depends on inverse powers of momenta, which for the \N=4 YM theory rules out all  loop UV divergences. Namely, the light-cone tree amplitude (\ref{Ampl}) has a factor \cite{Drummond:2008vq}
\be
{\cal P} (p_1,...p_n;  \eta_1, ..., \eta_n)_{YM}= {1\over {\cal N} (123...n)} \sum_{\{\alpha\}} R_{\alpha} (\lambda_i \tilde \lambda_i, \eta_i) \ .
\label{YM}\ee
Here $R_{\alpha} (\lambda_i \tilde \lambda_i, \eta_i)$ are dual superconformal invariants, they do nor eliminate the singularity in momenta of the light-cone amplitudes due to the term ${\cal N} ^{-1}(123...n)\equiv (\langle 12\rangle .... \langle n1 \rangle)^{-1}$. Neither does it the term $\left ( { p_{\bot l}\over (p_l ^{+})^{1/2}} \eta_l\right )^4$. For the 4-point amplitude at the l-loop order we can only multiply the tree-level expression in (\ref{YM}) by a function of Mandelstam variables of dimension zero. This means that the singularity cannot be removed from the light-cone amplitude and no local UV divergences are predicted in the light-cone supergraph computations at all-loop order  for the 4-point function.

In \N=8 SG the situation is similar with respect to the path integral derivation of the light-cone amplitude. The analog of N=4 YM structures is given in the paper.  The major difference is in evaluation of the non-polynomial in momentum space structure of the light-cone amplitude  due to a dimension-full coupling constant $\kappa^2$.

\section{Pure Gravity versus \N=8 SG on the Light Cone}

Here we present the steps which in the light-cone gauge in pure gravity will help to understand the situation. There is no supersymmetry!  Assume that we use the light-cone gauge for the gravitons and we have $h (x)$ and $\bar h(x)$ fields which are the only physical states out of all 10-component $h_{\mu\nu}$ gravitational fields in d=4. We use the gauge which eliminates 4 fields and the other 4 have algebraic equations of motion which can be used to replace them via some function of only $h (x)$ and $\bar h(x)$. This gives the Einstein action in the form where it depends only on $h (x)$ and $\bar h(x)$ and has some inverse powers of $\partial ^+$, but not inverse powers of  transverse directions.
\be
S_{cl}^{lc} [h, \bar h] ={1\over 2 \kappa^2}  \int d^4x ~h(x) \Box \bar h(x) +...
\label{action}\ee
Note that in this action the fields $h (x)$ and $\bar h(x)$ are unconstrained, so we can perform the computations of Feynman diagrams in pure gravity with just 2 propagating fields. In this form the fields $h (x)$ and $\bar h(x)$ have dimension zero, so that coupling terms can have any powers of $h (x)$ and $\bar h(x)$ with two derivatives and  all $\kappa$ dependence is outside the total action (like ${1\over g^2} F_{\mu\nu}^2$ in YM).

This is well known in YM case: if you integrate the light-cone superfield action you get the light-cone action for gluons etc. See for example the formula (3.13) for the light-cone N=4 SYM in components and the light-cone superfield action in eq. (4.10) of Brink, Lindgren, Nilsson paper in \cite{Brink:1982pd}.
In SYM the loop computations were performed in components, see for example the papers of Bassetto et al \cite{Bassetto:2007zj} and references therein. What they call ``transverse'' fields are exactly the components of the light-cone superfield for which they find all $Z$-factors are equal to 1.

So we would like to make a prediction for the light-cone pure gravity amplitudes which should come out from the light-cone Feynman graphs. It is convenient to switch to Fourier. We will make a  prediction for the outcome of the path integral
\be
Z= e^{i W[h_{in}, \bar h_{in}]}= \int dh \, d\bar h~   e^{i [S_{cl}(h, \bar h) +\int d^4 x  ( h_{in}(x) \Box \, \bar h(x)+cc)]}   .
\ee
Note that here the path integral depends on the on-shell free fields $h_{in}, \bar h_{in}$ with $\Box h_{in}= \Box \bar h_{in}=0$,  whereas the integration variables are the virtual fields $h,  \bar h$ which do not satisfy any equation. The propagator is $\Box ^{-1}$. The reason this is valid only in d=4 is that we have two propagating degrees of freedom for gravity only in d=4. In any other dimension graviton field $h_{\mu\nu}$ has a different number of physical degrees of freedom. Therefore our formalism will break unitarity in $d\neq 4$. Note that $W[h_{in}, \bar h_{in}] = \sum_{n=1}^{\infty} W^n[h_{in}, \bar h_{in}]$.

We claim that the 4-point part of the effective action  in d=4 in the light-cone pure gravity is given by
\be
 W^4[ h_{in},  {\bar h}_{in}]= \prod_{i=1}^{4}  \left (\int d^4p_i \delta (p_i^2) \right )  h_{in}(p_1) \,  h_{in}(p_2)  {\bar h}_{in}(p_3)  { \bar h}_{in}(p_4)  )~  \delta^4 \left(\sum_{k=1}^{k=4} p_k \right) {\cal A}_4^{lc}+
 (- - + +)+...
\label{funcGrav}\ee

Let us check the mass dimensions. Each integral $\int d^4 p h(p)\delta(p^2) $ has dimension zero as our space-time field $h(x)$, $ \delta^4 (\sum_{k=1}^{k=4} p_k )$ has dimension -4. The amplitude ${\cal A}_4^{lc}$ has to have dimension +4. The answer is of the form

\be
{\cal A}_4^{lc} (- - + +) = {1\over \kappa^2 stu} {( p_{\bot  1} p_2^+ -  p_{\bot  2} p^+_1)^4 \over (p_1^+ p_2^+)^{2}} {({\bar p}_{\bot 3} p_4^+ -  {\bar p}_{\bot 4}p^+_3)^4\over (p_3^+ p_4^+)^{2}} f(\kappa^2; s, t, u) \ .
\ee
Here $f(\kappa^2; s, t, u)$ is a function of dimension 0. In particular we can have $f^{tree}=\rm const$ and for the 3-loop UV divergence we have
\be
f^{3loop}(\kappa^2; s, t, u)= \kappa^6 stu \ln \Lambda
\ee
and
\be
{\cal A}_4^{lc} (- - + +) ^{3loop}= \kappa^4 {( p_{\bot 1}  p_2^+ -  p_{\bot 2} p^+_1)^4 \over (p_1^+ p_2^+)^{2}} {( {\bar p}_{\bot 3} p_4^+ -  {\bar p}_{\bot 4} p^+_3)^4\over (p_3^+ p_4^+)^{2}} \ln \Lambda \ .
\ee
The 4-graviton amplitude is local in transverse directions (Fourier of $p_{\bot},  \bar p_{\bot}$, the only non-locality is in direction $x^-$ (and $x^+$ when the improved $1/p^+$ is used).

 Note that the amplitude here is defined as a coefficient in front of the second power of the $h_{in}$ and second power of the $\bar h_{in}$ field, since we compute the path integral integrating the fields $h$, $\bar h$ out.

We may have computed the same path integral in any Lorentz covariant  gauge, e.g. in de Donder gauge  $D^\mu h_{\mu\nu}-{1\over 2} D_\nu h =0$ but introducing only the sources to physical states. From the equivalence theorem it follows that the amplitude must be the same: namely we can rewrite
\be
{\cal A}_4^{cov} (- - + +) = {1\over \kappa^2 stu} \langle 12\rangle ^4 [34]^4
 f(\kappa^2, s, t, u)
\ee
and ${\cal A}_4^{lc} (- - + +) ^{3loop} ={\cal A}_4^{cov} (- - + +)^{3loop}$, where
\be
{\cal A}_4^{lc} (- - + +) ^{3loop}= \kappa^4 {( p_{\bot 1}  p_2^+ -  p_{\bot 2} p^+_1)^4 \over (p_1^+ p_2^+)^{2}} {( {\bar p}_{\bot 3} p_4^+ -  {\bar p}_{\bot 4}p^+_3)^4\over (p_3^+ p_4^+)^{2}} \ln \Lambda= \kappa^4   \langle 12\rangle ^4 [34]^4  \ln \Lambda \ .
\ee
{\it Nothing forbids the 3-loop log divergence when computing the amplitudes in the light-cone gauge in pure gravity.}

 Now let us see in the same setting what is going on in \N=8 when we use the light-cone superfields. We must use the Mandelstam-type action (\ref{action2}) where only 8 $\theta$'s are manifest in the action (4 in  \N=4 SYM).
We compute the path integral where the propagating field is a single chiral scalar unconstrained superfield $\phi(x, \theta)$ and we introduce a source $ \phi_{in}(x, \theta)$ which is a free field $\Box \phi_{in}(x, \theta)=0$
\be
 \left(e^{i W[\phi_{in}]}\right)_{{\cal N}=8} = \int d\phi  e^{i (S_{cl}[ \phi(x, \theta)] +\int d^4 x d^8 \theta  \phi_{in}(x, \theta)\vec {{\Box}} \phi(x, \theta))} \ .
\label{pathint1}\ee
{\it Note now that the 4-point amplitude is a coefficient in front of 4th power in  superfields $ \phi_{in}(x, \theta)$. Each has a helicity -2, therefore the amplitude will have helicity +2 at each point.} Fourier moves 4+8 $(x, \theta)$ into 4+8 $(p, \eta)$.
The analog of (\ref{funcGrav}) is
\be
 W^4[\Phi_{in}]= \prod_{i=1}^{4}  \left (\int d^4p_i \delta (p_i^2) d^8 \eta_i \,  \Phi _{in}(p_i, \eta_i)\right ) \, \delta^4 (\sum_{k=1}^{k=4} p_k ) \delta^8 \left (\sum _{l=1} ^{l=4} (p_l ^{+})^{1/2} \eta_l\right ) {\cal A}_n^{lc}
 (p_1,...p_4) \ .
\label{func1a}\ee
All fields are of the same nature since we have a scalar chiral superfield (as different from gravity where you have $h$ and $\bar h$, here we have only $\Phi$. Secondly, we have an $4+8$ delta function from Fourier,
 $ \delta^4 (\sum_{k=1}^{k=4} p_k ) \delta^8 \left (\sum _{l=1} ^{l=4} (p_l ^{+})^{1/2} \eta_l\right )$. The $4+8$ delta functions have dimension $-4+4=0$ ($\eta$ has dimension zero). Thus we have a dimension zero light-cone amplitude (in pure gravity it was dimension +4). The answer is
\be
\left({\cal A}_4^{lc}
 (p_1,...p_4;  \eta_1, ..., \eta_4)\right)_{{\cal N}=8}= \left ( \sum _{l=1} ^{l=4} { p_{\bot l}\over (p_l ^{+})^{1/2}} \eta_l\right )^8  {\cal P} (+ + + +) \ .
\label{A4b}\ee
Dimension of  $\left ( \sum _{l=1} ^{l=4} { p_{\bot l} \over (p_l ^{+})^{1/2}} \eta_l\right )^8$ is $+4$, therefore the dimension of ${\cal P} (+ + + +)$ is $-4$. In analogy with what we have done in pure gravity we find
\be
{\cal P}^{0}(++++)=i   {1\over \kappa^2 stu}{[ij ]^4\over \langle i'j'\rangle^4 } f(\kappa^2, s,t,u) \ ,
\ee
where $i, j$ and $i', j'$ are arbitrary 4 different points.  Note that  in light-cone gauge
\be
\langle k l \rangle \equiv { p_{\bot k}  p_l^+ -  p_{\bot l} p^+_k\over (p_k^+ p_l^+)^{1/2}} \ ,  \qquad  [ k l ] \equiv { {\bar p}_{\bot k} p_l^+ - {\bar  p}_{\bot l} p^+_k\over (p_k^+ p_l^+)^{1/2}} \ .
\label{angularsquare}\ee
For the tree case $f=\rm const$ and for the 3-loop divergence $f= \kappa^6 stu \ln \Lambda$. This will give us the 3-loop UV divergent amplitude for superfields in the form
\be
{\cal A}^{lc}_{UV3loop}= \kappa^4 \left ( \sum _{l=1} ^{l=4} { p_{\bot l}\over (p_l ^{+})^{1/2}} \eta_l\right )^8 \left ({ {\bar p}_{\bot 3} p_4^+ - {\bar  p}_{\bot 4} p^+_3\over  p_{\bot 1} p_2^+ -  p_{\bot 2} p^+_1}\right )^4 \left ({p^+_1 p^+_2\over p^+_3 p^+_4}\right)^2  ~
 \ln \Lambda \ .
\ee
This is the answer you would get when performing the supergraph Feynman rules starting with the action (\ref{action2}) where the propagating fields are the chiral scalar superfields $\phi (x, \theta)$. But we cannot have a non-locality in the transverse directions, for the log divergent terms, like the one related to the inverse power of $ p_{\bot 1} p_2^+ -  p_{\bot 2} p^+_1$. This eliminates the 3-loop log in \N=8.

The reason why this is the correct answer is explained at the end of the subsection  4.1 where we explain the relation to the square of the Bel-Robinson tensor:  our light-cone 4-superfield amplitude will give a correct 4-graviton amplitude after $4\times 8$ $\eta$-integrations are performed. However, the divergent part of the amplitude cannot appear in the underlying  light-cone supergraph computations.

\section{Mandelstam-Leibbrandt  Prescription}

The locality/non-locality issue in the light-cone superspace has certain issues which were known to the experts from the beginning. Namely,  one has to use an  improved prescription and $p^+$. This leads to a  non-locality in the $x^-$-direction as well as in  $x^+$.

The  Mandelstam-Leibbrandt   prescription \cite{Mandelstam:1982cb}, \cite{Leibbrandt:1983pj} for dealing
with $1/p_+$ factors inside Feynman integrals involves $p_-$ in the following way:
\be
\left ({1\over p^+}\right)_{ML} = {1\over p^+  + i\epsilon \; sign \; p^- }
\label{ML}\ee
with $\epsilon \rightarrow 0$. In other words,  it only involves $p^-$- component, not the transverse ones related to $x^1, x^2$. Let us look at this in detail.

It is explained in \cite{Mandelstam:1982cb},  that this $\epsilon$-prescription allows to continue the integrals to an Euclidean space-time. Without $\epsilon$-prescription this was not possible, there were poles in $p^+$,  and therefore the ``naive'' power counting of UV divergences was not valid. But with improved prescription one can go around the poles and therefore the Euclidean continuation is possible.

The naive definition of the inverse of $p^+$ was
\be
{1\over p^+} f(x^-)= -{1\over2} i \int dx^{-'}\epsilon( x^{-'}- x^-) f(x^{-'}) \ .
\ee
The  improved prescription replaces the ${1\over 2} \epsilon(x^-)$ by the following expression
\be
\epsilon (p^-)\theta(x^-) - \epsilon (-p^-)\theta(-x^-) \ .
\ee
In coordinate space this means that $\epsilon(p^-)  $ becomes
\be
{1\over2} i \int dx^{+'}\epsilon( x^{+'}- x^+) \ .
\ee

Thus, if we would perform the computations with improved prescription, we may see non-locality in directions $x^-$ as well as $x^+$. But not in $x^1$ or  $x^2$ or in the chiral directions $x^1\pm i x^2$. This means that if the supergraph computations produce a non-locality in the transverse directions, such terms cannot  support  logarithmic divergences.

\end{appendix}

\end{document}